\documentclass[sigconf]{acmart}
\AtBeginDocument{%
  \providecommand\BibTeX{{%
    \normalfont B\kern-0.5em{\scshape i\kern-0.25em b}\kern-0.8em\TeX}}}

\usepackage{hyperref}
\usepackage{color}
\usepackage{makecell}
\usepackage{url}
\usepackage{balance}
\usepackage{tabularx}

\usepackage{bm}
\usepackage{pifont}
\newcommand{\xmark}{\ding{55}}

\usepackage{algorithmic}
\usepackage{graphicx}
\usepackage{subcaption}
\usepackage[boxed,ruled,lined]{algorithm2e}
\usepackage{amsthm}
\usepackage[american]{babel}
\usepackage{setspace}
\usepackage{enumerate}
\usepackage{multirow}
\usepackage{url}

\newcommand{\argmax}{\operatornamewithlimits{arg\,max}}
\newcommand{\argmin}{\operatornamewithlimits{arg\,min}}

\setcopyright{acmcopyright}
\copyrightyear{2018}
\acmYear{2018}
\acmDOI{XXXXXXX.XXXXXXX}
\acmConference[Conference acronym 'XX]{Make sure to enter the correct conference title from your rights confirmation emai}{June 03--05,
  2018}{Woodstock, NY}
\acmPrice{15.00}
\acmISBN{978-1-4503-XXXX-X/18/06}

\begin{document}

\title{HyperBandit: Contextual Bandit with Hypernewtork  
for Time-Varying User Preferences in Streaming Recommendation}
\def\shorttitle{HyperBandit} 

\author{Chenglei Shen}
\affiliation{%
  \institution{Gaoling School of Artificial Intelligence\\Renmin University of China}
  \city{Beijing}
  \country{China}}
\email{chengleishen9@ruc.edu.cn}

\author{Xiao Zhang}
\authornote{Xiao Zhang is the corresponding author. The work was partially done at Beijing KeyLaboratory of Big Data Management and Analysis Methods.}
\affiliation{%
  \institution{Gaoling School of Artificial Intelligence\\Renmin University of China}
  \city{Beijing}
  \country{China}}
\email{zhangx89@ruc.edu.cn}

\author{Wei Wei}
\affiliation{%
  \institution{CCIIP Laboratory\\Huazhong University of Science and Technology}
  \city{Wuhan}
  \country{China}}
\email{weiw@hust.edu.cn}

\author{Jun Xu}
\affiliation{%
  \institution{Gaoling School of Artificial Intelligence\\Renmin University of China}
  \city{Beijing}
  \country{China}}
\email{junxu@ruc.edu.cn}

\renewcommand{\shortauthors}{Chenglei Shen et al.}

\begin{abstract}
In real-world streaming recommender systems, user preferences often dynamically change over time (e.g., a user may have different preferences during weekdays and weekends). Existing bandit-based streaming recommendation models only consider time as a timestamp, without explicitly modeling the relationship between time variables and time-varying user preferences. This leads to recommendation models that cannot quickly adapt to dynamic scenarios. To address this issue, we propose a contextual bandit approach using hypernetwork, called HyperBandit, which takes time features as input and dynamically adjusts the recommendation model for time-varying user preferences. 
Specifically, HyperBandit maintains a neural network capable of generating the parameters for estimating time-varying rewards, taking into account the correlation between time features and user preferences. 
Using the estimated time-varying rewards, a bandit policy is employed to make online recommendations by learning the latent item contexts. To meet the real-time requirements in streaming recommendation scenarios, we have verified the existence of a low-rank structure in the parameter matrix and utilize low-rank factorization for efficient training. 
Theoretically, we demonstrate a sublinear regret upper bound against the best policy. Extensive experiments on real-world datasets show that the proposed HyperBandit consistently outperforms the state-of-the-art baselines in terms of accumulated rewards.

\end{abstract}


\maketitle
\section{Introduction}
\label{sec:intro}



While the demand for personalized recommendations has increased due to the growth of online platforms and user-generated content, it is crucial to emphasize that the recommendation models need to be updated frequently and integrated with online recommender systems to ensure optimal performance in real-time. This makes streaming recommendation a highly active area of research aimed at continuously updating the model based on users' latest interactions with the platform and delivering relevant and timely suggestions to users \cite{Chandramouli2011StreamRec, Chang2017Streaming, Jakomin2020Simultaneous, Wang2018Neural,Zhang2022Counteracting, Wang2018Streaming}.


\begin{figure}[!t]
\centering
\begin{subfigure}[b]{0.22\textwidth}
\includegraphics[width=\textwidth]{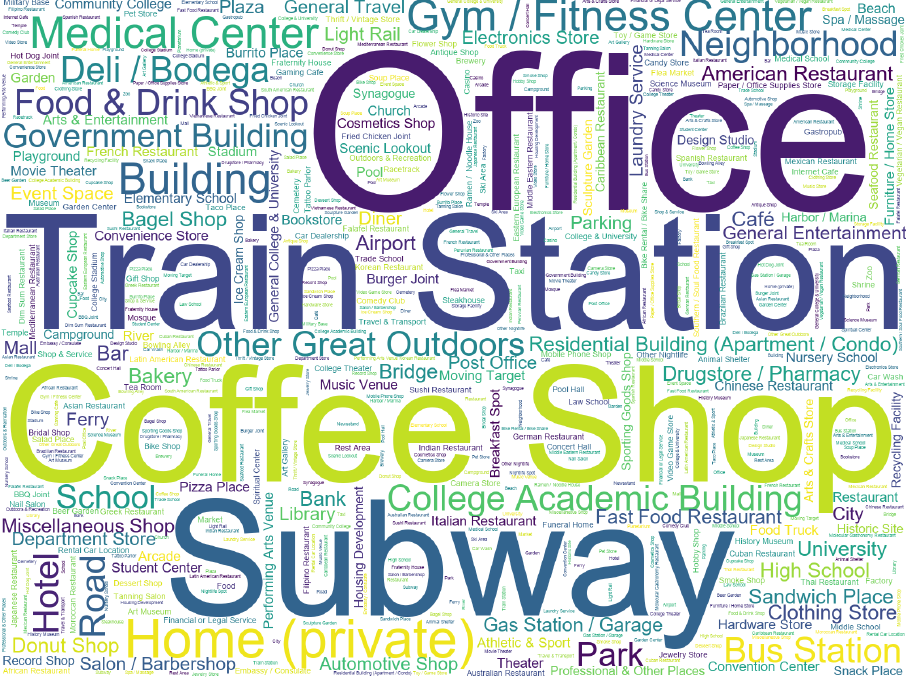}
\caption{\texttt{weekday morning}}
\end{subfigure}
\hfill
\begin{subfigure}[b]{0.22\textwidth}
\includegraphics[width=\textwidth]{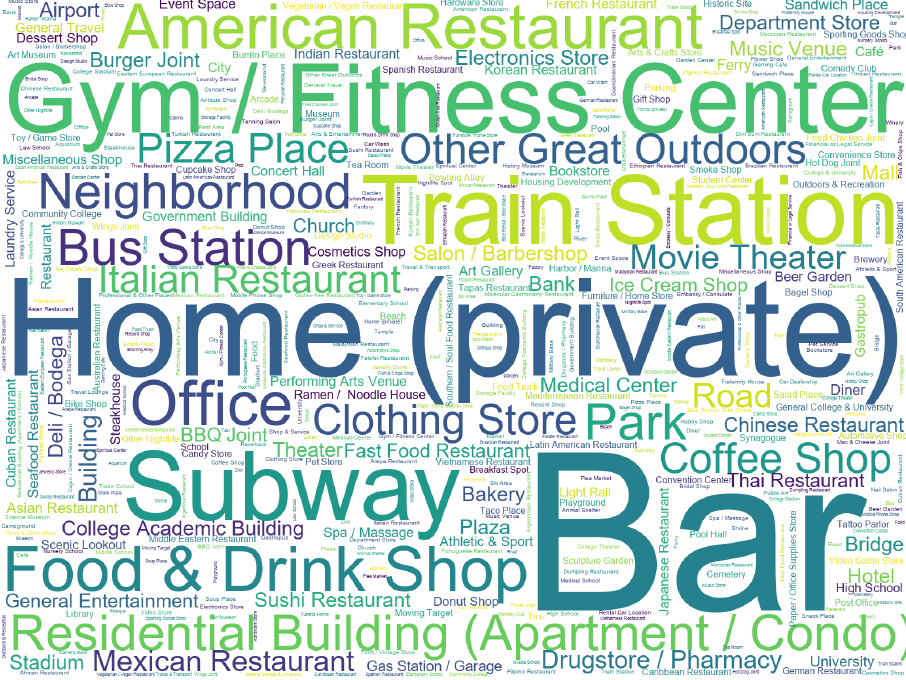}
\caption{\texttt{weekday night}}
\end{subfigure}

\begin{subfigure}[b]{0.22\textwidth}
\includegraphics[width=\textwidth]{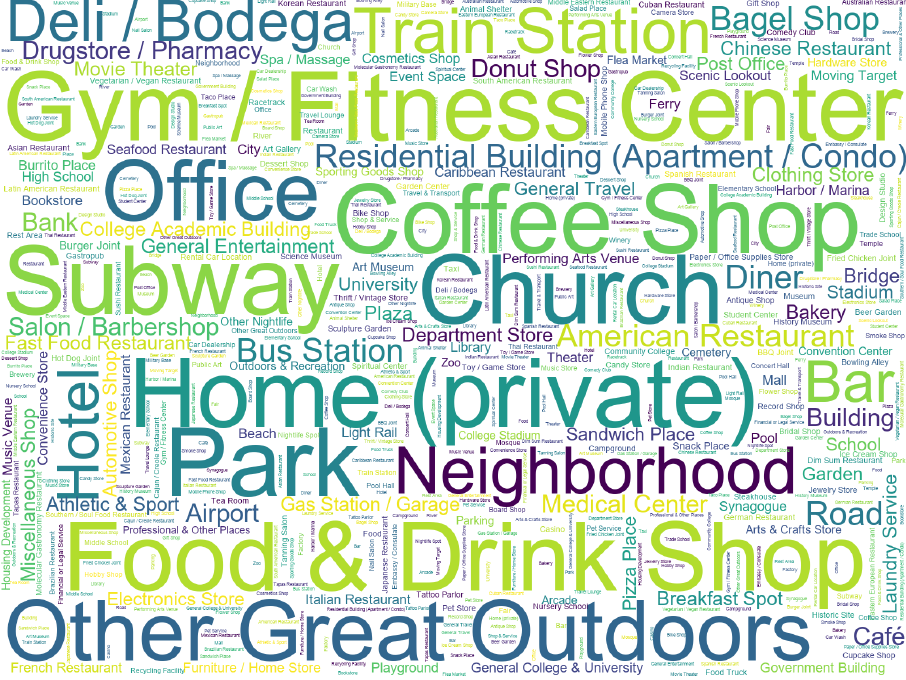}
\caption{\texttt{weekend morning}}
\end{subfigure}
\hfill
\begin{subfigure}[b]{0.22\textwidth}
\includegraphics[width=\textwidth]{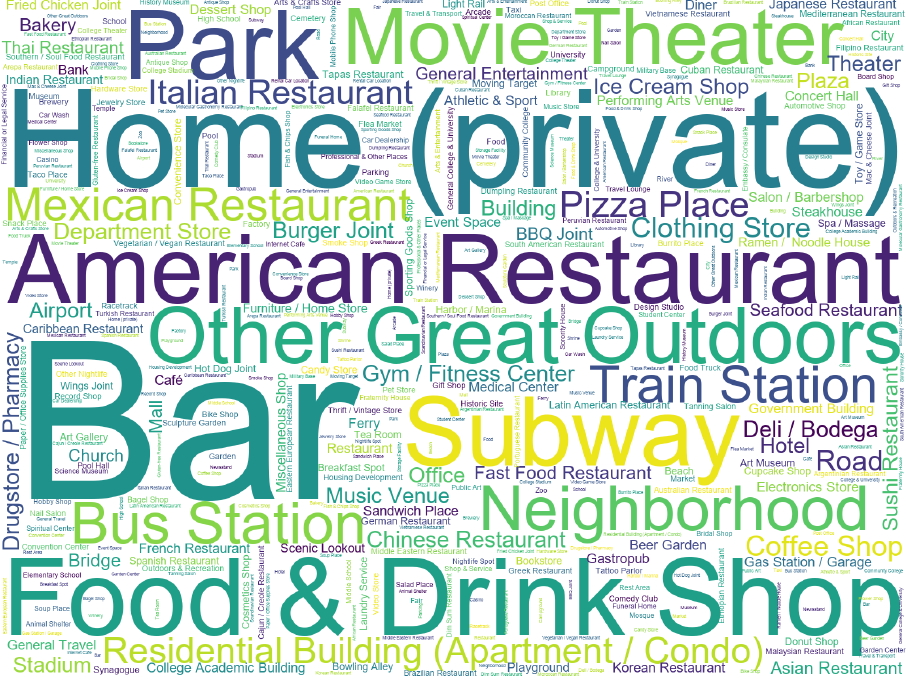}
\caption{\texttt{weekend night}}
\end{subfigure}
\vspace{-1ex}
\caption{The illustrations of the periodic shift on points-of-interest (POI) dataset Foursquare-NYC, representing word clouds of POI for morning/night on weekdays/weekends.}
\label{fig:NYCtimevarying}
\vspace{-3ex}
\end{figure}

Nonetheless, streaming recommendation confronts a significant challenge in the form of the phenomenon of time-varying user preferences~\cite{ditzler2015learning}. Users' preferences change dynamically over time due to several factors such as seasonality, holidays, or circadian rhythm. As illustrated in Fig.~\ref{fig:NYCtimevarying}, users tend to check in at places such as ``Office'' and ``Coffee Shop'' on weekday mornings, while at places like ``Gym / Fitness Center'' and ``Church'' on weekend mornings, demonstrating a weekly periodicity.  
In contrast to morning preferences, users tend to visit bars and spend time at home during evening hours regardless of whether it is a weekday or weekend, indicating a daily periodicity. 
Another interesting example of short video recommendation is that users exhibit a tendency to watch cartoons specifically on weekends,  while preferring other types of content on weekdays. 
These recurring patterns highlight the importance of considering time-varying user preferences to avoid sub-optimal recommendations. 
Consequently, devising effective and efficient approaches to address the issue of users' periodic time-varying preference is critical for achieving high-quality streaming recommendation.



As a classic framework for online learning, multi-armed bandit (MAB) algorithms have gained significant attention in recent years. A variation of MAB, known as contextual bandits \cite{langford2007epoch, li2010contextual,wu2018learning}, has achieved considerable success in various online services by utilizing both user feedback and contextual information related to users and items, which make it particularly advantageous in  streaming recommendation scenarios. Most existing contextual bandit algorithms are constructed under stationary environment, i.e. users’ preferences remain static over time \cite{li2010contextual,Balseiro2019Contextual,han2020sequential}. However, the environment is always non-stationary in reality indicating time-varying user preferences. Some studies have noticed this problem and relaxed the assumption to the piecewise stationary environment \cite{wu2018learning,xu2020contextual}, which enables algorithms to adaptively detect user preferences change points and discard learned model parameters for relearning. 
These approaches may result in performance fluctuations when handling periodic changes in user preferences. The primary reason is that these algorithms fail to recognize the periodic information of user preferences in an online manner and retrain the model even if the current period has occurred in the past. Currently, there is a notable research gap in the domain of streaming recommendations in periodic environments.





In this paper, we focus on a realistic environment setting where the reward function (i.e., the generation mechanism of user feedback) exhibits periodicity over time. Specifically, a large time period can be divided into multiple smaller periods in a periodic manner (e.g., based on the specific day of the week and different time slots within a day), and the reward function demonstrates a similar distribution whenever the same time period is encountered. Moreover, these time periods can be observed by the model and utilized for periodicity modeling and online adjustment of its user preference module in various streaming recommendation scenarios.

As a specific solution to the aforementioned process, we propose a novel contextual bandit algorithm called HyperBandit, which consists of two levels of model structures:
1). A \emph{bandit policy} is designed to learn the latent features of items in an online fashion and combine them with the user preference matrix to execute online recommendations with effective exploration. 
2). A \emph{hypernetwork} takes the information of the time period as inputs and generates the parameters of the user preference matrix in the bandit policy. This hypernetwork captures the periodicity of user preferences over time and enables efficient online updating through low-rank factorization. 
The contributions are summarized as follows:
\begin{enumerate}[(1)]
    \item We propose a novel contextual bandit algorithm called HyperBandit, along with an efficient online training method utilizing low-rank factorization. HyperBandit explicitly models the periodic variations in user preferences and dynamically adjusts the recommendation policy based on time features.
    \item We provide a sublinear regret guarantee for HyperBandit, ensuring the convergence of the online learning process. Additionally, we conduct empirical analysis on the low-rank structure in the parameter matrix, validating the rationale behind training via low-rank factorization.
    \item We perform extensive experiments on various streaming recommendation tasks, including the recommendation of short videos and points of interest (POI), demonstrating the efficiency and effectiveness of the proposed HyperBandit algorithm.
\end{enumerate}


\section{Related Work}
{\bf Hypernetworks (HNs)}  have  been introduced by Ha et al. \cite{ha2016hypernetworks}, drawing inspiration from the genotype-phenotype relation in cellular biology. HNs present an approach of using one network (hypernetwork) to generate weights for a second network (target network). In recent years, HNs are widely used in various domains such as computer vision \cite{klocek2019hypernetwork}, language modeling~\cite{suarez2017language}, sequence decoding~\cite{nachmani2019hyper}, continual learning ~\cite{von2019continual}, federated learning~\cite{shamsian2021personalized}, multi-objective optimization \cite{Navon2021Learning,Chen2023Controllable}, and hyperparameter optimization~\cite{mackay2019self}. Navon et al.~\cite{Navon2021Learning} proposed a unified model to learn the Pareto front based on HNs that can be applied to a specific objective preference at inference time. 
von Oswald et al.~\cite{von2019continual} presented a task-aware method for continual model-based reinforcement learning using HNs, which allows the entire network to change between tasks as well as retaining performance on previous tasks. HNs have been widely applied in offline learning, but there is a lack of research on how to enhance the controllability of models through hypernetworks in online learning and streaming applications.

{\bf Bandits in non-stationary environment}
 have attracted extensive attention in both theory and applications in recent years. One common setting for non-stationary environments is the abruptly changing or piecewise-stationary environment, where the environment undergoes sudden changes at unknown time points while remaining stationary between consecutive change points. Under the piecewise-stationary assumption, the problem has been well studied in the classical context-free setting \cite{hartland2006multi,garivier2008upper,yu2009piecewise,slivkins2008adapting}. 
 Yu et al.~\cite{yu2009piecewise} proposed a windowed mean-shift detection algorithm to identify potential abrupt changes in the environment. They provided an upper bound on regret of $O\left(\Gamma_T \log (T)\right)$ for their algorithm, where $\Gamma_T$ represents the number of ground-truth changes up to time $T$.
 Within the contextual bandit setting, limited attention has been given to addressing non-stationary environments~\cite{hariri2015adapting,wu2018learning,xu2020contextual}. 
 Wu et al.~\cite{wu2018learning} developed a hierarchical bandit algorithm capable of detecting and adapting to changes by maintaining multiple contextual bandits. 
More recently, Xu et al.~\cite{xu2020contextual} addressed the challenge of time-varying preferences by employing a change-detection procedure to identify potential changes on the preference vectors. However, little attention has been given to addressing the issue of periodic reward drift that this paper focuses on. 



\section{Problem Formulation}

\subsection{\mbox{Bandit-based Streaming Recommendation}}

Streaming recommendation can be formulated as a problem of sequential decision making, where the online service platform recommends the most relevant item $a \in \mathcal{A}$ (such as videos, music, or POIs) to a user $u \in \mathcal{U}$ in an online manner. 
Contextual bandit algorithms are well-suited for addressing streaming recommendation problems.
More specifically, the candidate item set $\mathcal{A}$ could be viewed as the action space  of the bandit algorithm, while the context space $\mathcal{S}$ summarizes the feature information of users and items,  where each item $a$ and user $u$ can be associated with context feature vectors denoted by $\bm c_a$ and $\bm c_u$, respectively. 
At time step $t$, given a subset of the action space $\mathcal{A}_t \subseteq \mathcal{A}$ and a user, an item is selected by a recommendation policy and recommended to the user.  
After one item is recommended, the item may be clicked by the user (i.e, positive user feedback) or skipped (i.e, negative user feedback). Thereafter the true reward defined on the user feedback is received and could be used for updating the current recommendation policy, which will be adopted for the next recommendation.

The above process can be formalized as a contextual bandit problem for streaming recommendation, and represented using a 4-tuple $\left\langle\mathcal{A}, \mathcal{S}, \pi, r \right\rangle$:



\textbf{Action space $\mathcal{A}$} denotes a given candidate action set, where each action (also called arm) corresponds to a specified candidate item. At each time step, a dynamic action space is selected as the candidate item set for recommendation.
That is, at time step $t$, a candidate item set $\mathcal{A}_t \subseteq \mathcal{A}$ is recalled by some strategy, and choosing an action $a_{I_{t}}$ from $\mathcal{A}_{t}$ means that the corresponding item is recommended to the user, where $I_{t} \in |\mathcal{A}_t|$ denotes the index of the recommended item at time $t$. 


\textbf{Context space $\mathcal{S}$} summarizes the context feature information of users and items, denoted by $\bm c_u \in \mathbb{R}^{d_u}$ and $\bm c_a \in \mathbb{R}^{d_a}$, respectively. 
In this paper, in particular, we consider splitting the item context $\bm c_a \in \mathbb{R}^{d_a}$ into two parts: the observed features $\bm s_{a}\in \mathbb{R}^{o_a}$, and the latent features $\bm x_{a} \in \mathbb{R}^{l_a}$ that needs to be learned. Here, $\bm c_a = [\bm s_a^\intercal, \bm x_a(t)^\intercal]^\intercal$ and the dimension of $\bm c_a$ is given by $d_a = o_a + l_a$.

\textbf{Policy $\pi : \mathcal{S} \rightarrow \mathcal{A}$} describes the decision-making rule of an agent (i.e., the recommendation model), which selects an action for execution according to the relevance score of each action. at time $t$, given a candidate item set $\mathcal{A}_{t}$ and user $u \in \mathcal{U}$, a \emph{relevance score function} $f_t$ treats context features of user and item  in context space 
(i.e., $\bm c_{u}$ and $\bm c_a$ ) 
as inputs and determines which action to take: $a_{I_t}:=\arg \max _{a \in \mathcal{A}_{t}} f_t\left(\bm c_{u} , \bm c_a \right)$.

\textbf{Reward $r$} is defined upon the user feedback. Specifically, at time~$t$, after recommending the item $a_{I_t} \in \mathcal{A}_t$ to a user $u$, a corresponding reward $r (u, a_{I_t}) \in\{0,1\}$ is observed, which implicitly indicates whether the user feedback is negative or positive to the item $a_{I_t}$. 
However, the feedbacks from the same user towards the same item at different time may be quite different, which means that time-varying user preferences exist in it.

Table~\ref{tab:notations} summarizes the notations used throughout the paper.

\begin{table}[t]
    \centering
    \footnotesize
    \renewcommand{\arraystretch}{1.2} 
        \caption{A summary of notations.}
        \vspace{-1ex}
        \begin{tabularx}{0.45\textwidth}{|c|X|}
        \hline  
        {\bf Symbol} & {\bf Explanation}\\
        \hline  
        $ [n]$ & $[n]: = [1,2, \ldots, n]$
        \\
        \hline
        $t$& Time step $t \in [T]$\\
        \hline
        $ \mathcal{A}$ & Action space, i.e., the candidate item set 
        \\
        \hline
        $|\mathcal{A}|$& The cardinality of set $\mathcal{A}$ 
        \\
        \hline 
        $\bm c_{u} \in \mathbb{R}^{d_u} $& Context feature vector of a user $u$\\
        \hline
        $\bm c_{a} \in \mathbb{R}^{d_a} $& Context feature vector of a candidate item $a$\\
        \hline
        $\bm s_{a}\in \mathbb{R}^{o_a}$ &  Observed features of a candidate item $a$\\
        \hline    
        $\bm x_{a} \in \mathbb{R}^{l_a}$& Latent features of a candidate item $a$\\ 
        \hline  
        \multirow{2}{*}{$p \in \mathcal{P}$ } &  Time period variable, takes values in the range $\mathcal{P} := \{0, 1, \dots, 34\}$, representing the 35 time periods within a week\\
        \hline  
        $\bm s_{p} \in \mathbb{R}^{d_p}$& Time period embedding of time period $p$\\ 
        \hline  
        
        $ \bm \Theta_p^* $& True user preference matrix at the time period $p$\\
        \hline  
        \end{tabularx}
        \label{tab:notations}
\end{table}

\subsection{Time-Varying User Preferences}

In this section, we formally describe the time-varying user preferences mentioned in the introduction. 

As defined in Table~\ref{tab:notations}, we first introduce the time period variable $p$ to measure specific temporal patterns, including hours of the day and different days of the week. Specifically, we divide a week into seven days, from Monday to Sunday, and further divide each day into the following five sessions: the morning (8:00 AM to 11:30 AM), the noon (11:30 AM to 2:00 PM), the afternoon (2:00 PM to 5:30 PM), the night (5:30 PM to 10:00 PM), and the remaining period. 
Then, the time period variable, $p$, encompasses 35 distinct values spanning from 0 to 34 in a sequential order. Each time period can be encoded to derive its respective \emph{time period embedding}, denoted as $\bm s_p \in \mathbb{R}^{d_p}$.


Under the traditional assumption of a stationary environment, the mechanism of user feedback should be consistent at every time period.
That is, the \emph{reward generation probability}, represented as $\operatorname{Pr} \left\{r (u, a) =1 \mid \bm c_u, \bm c_a\right\}$, is assumed to remain constant across all time step $t \in [T]$. This implies that the level of preference that user $u$ has for the recommended item $a$ is independent of the specific time at which the recommendation is made. 
However, in real-world streaming recommender systems, users' preferences change with time periodically, which has been observed in \cite{gao2013exploring}. For example, users usually visit office at weekday morning and bars at night. That is, the user feedback towards office may be different at different time period. In other words, given the context $\bm c_u, \bm c_a$, the current time period $p$ and the corresponding time period embedding $\bm s_p$, the following inequality may hold:
\begin{equation}
\operatorname{Pr}\left\{r(u, a)=1 \mid \bm c_u, \bm c_a, \bm s_{p=i}\right\} \neq \operatorname{Pr}\left\{r(u, a)=1 \mid \bm c_u, \bm c_a, \bm s_{p=j}\right\},
\end{equation}
where $i\neq j$ and $i,j \in \left\{ 0,\dots,34\right\}$, and $\operatorname{Pr}\left\{r=1 \mid \bm c_u, \bm c_a, \bm s_{p}\right\}$ denotes the \emph{time-varying reward generation probability} indicating how much the user $u$ prefers the recommended item $a$ at time period $p$.

Formally, we can represent the observed reward generated by the time-varying reward generation probability as $r(u, a, p)$. 
Given a user $u$, a item $a$, and a time period $p$, the generation process of the observed reward can be formalized as $r (u, a, p)  := r^* (u, a, p) + \eta$, where $r^* (u, a, p)$ represents the \emph{true reward}, and $\eta$ is a random variable drawn from a distribution with zero mean. This additional error term $\eta$ captures the noise or uncertainty present in the observations.
Clearly, we have the following expected reward: $$\mathbb{E} [r (u, a, p)] = r^* (u, a, p) = \operatorname{Pr}\left\{r (u, a, p)=1 \mid \bm c_u, \bm c_a, \bm s_{p}\right\}.$$

Next, we make specific assumptions about the form of the expected reward $\mathbb{E} [r (u, a, p)]$, i.e., the true reward. One straightforward approach is to concatenate the time period embedding with the context feature vectors. However, existing research~\cite{galanti2020modularity} has shown that directly concatenating features from different spaces can make it difficult to capture meaningful information (i.e., time-varying information in this paper). 
To address this issue, we extend the existing linear expected reward in the contextual bandit setting by introducing the true user preference matrix $\bm \Theta_p^*$.
Then, we can specify the true reward as the following \emph{time-varying true reward}:
\begin{equation}
\label{eq:hb:time_reward}
r^*(u, a, p) := \bm c_a^\intercal \bm \Theta_p^* \bm{c}_u, 
\end{equation}
where the \emph{true user preference matrix} $\bm \Theta_p^* \in \mathbb{R}^{d_a \times d_u}$ is utilized to map the user contexts through a linear mapping, taking into account the time period $p$ as well as its embedding $\bm s_p$ as conditions. By applying this mapping, the resulting vector $\bm \Theta_p^* \bm c_u$ can effectively capture the time-varying user preferences across different time periods.
In particular, when $d_a = d_u$ and $\bm \Theta_p^*$ is the identity matrix, the time-varying true reward degenerates to a fixed true reward in the traditional contextual bandit setting.

\section{\mbox{HyperBandit: The Proposed Algorithm}}
We propose a novel bandit algorithm tailored for periodic non-stationary streaming recommendation, called HyperBandit. 


\subsection{Algorithm Overview}

\begin{figure}[!t]
    \centering
    \includegraphics[width=\linewidth]{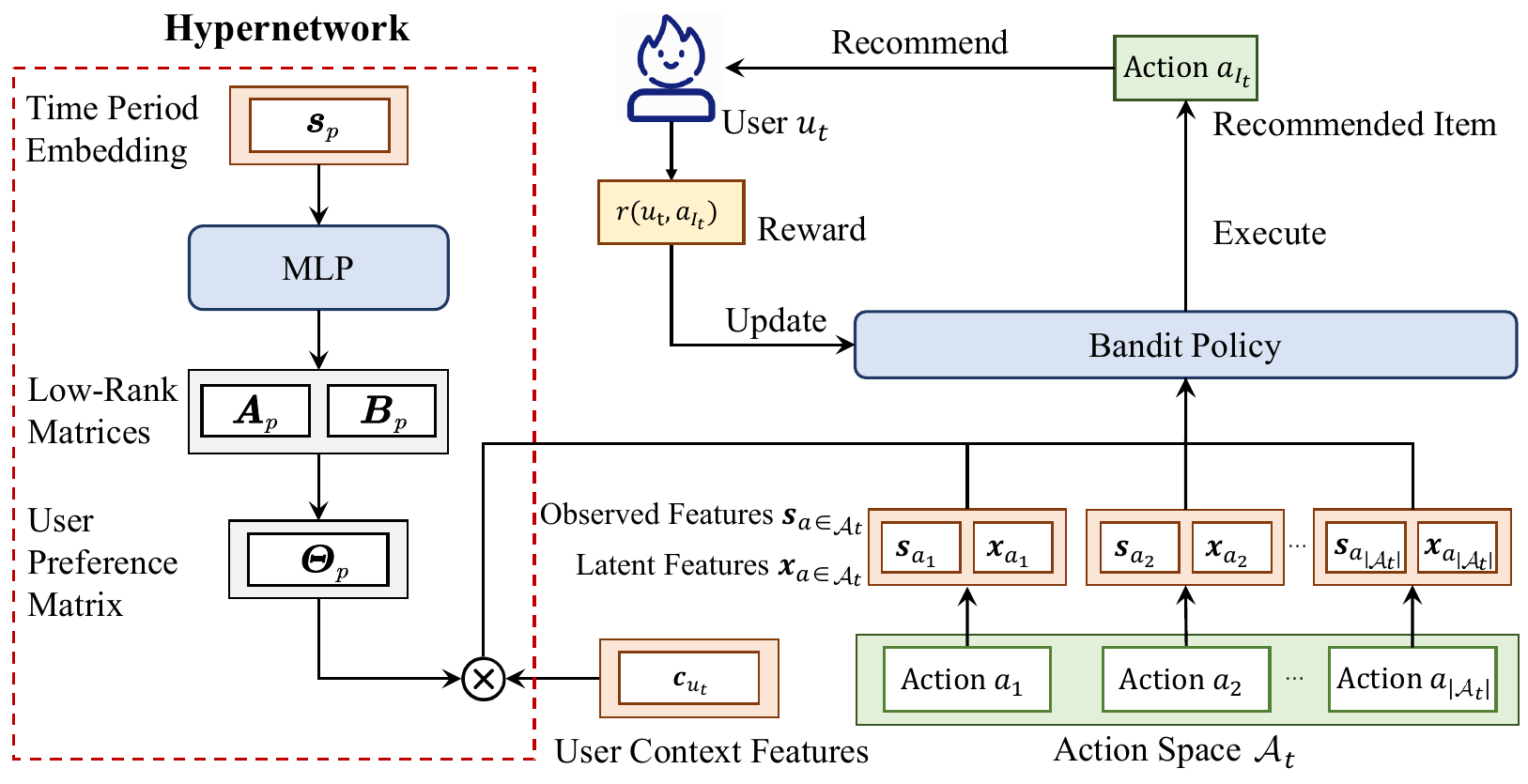}
    \caption{The structure of HyperBandit at time $t$.}
    \label{fig:hyperbandit_structure}
\end{figure}



Fig.~\ref{fig:hyperbandit_structure} illustrates the structure of HyperBandit. Given the current time period as input, a hypernetwork generates a user preference matrix that maps user context features to a time-aware preference space. The mapped features, along with the item context features, are then utilized by the bandit policy to recommend a suitable item to the current user.


HyperBandit consists of the following two components. Firstly, it utilizes a bandit policy to update the latent features of items at each time step, enabling the preservation of time-varying latent features to capture distribution shifting. 
Secondly, it employs a hypernetwork that is trained in a mini-batch manner to adaptively adjust the user preference matrix in the bandit policy for a given time period. The algorithm's detailed procedure is outlined in Algorithm \ref{alg:hyperbandit}. It's important to note that the user preference matrix is estimated using two low-rank matrices during training. This specific training technique will be discussed in detail in Sec.~\ref{sec:efficienttraining}.



\subsection{Hypernetwork Assisted Bandit Policy}

\subsubsection{Bandit Policy using User Preference Matrix}
To estimate the time-varying true
reward and account for the user preference shift in each time period, we propose a novel bandit policy that utilizes an estimate of the true user preference matrix $\bm \Theta_p^*$ in Eq.~\eqref{eq:hb:time_reward}. Specifically, the \emph{estimated user preference matrix}, denoted as $\bm \Theta_p \in \mathbb{R}^{d_a\times d_u}$, captures the changes in user preferences during time period $p$. We estimate $\bm \Theta_p$ using a hypernetwork, which will be introduced in Sec.~\ref{sec:HB:Hypernetwork}. 
The estimated user preference matrix allows us to adapt our bandit policy to the evolving user preferences. 
Formally, 
assuming that time step $t$ belongs to time period $p$, 
given a user context $\bm c_u \in \mathbb{R}^{d_u}$ and the estimated user preference matrix, 
the following ridge regression over the current interaction history is employed to estimate the item context $\bm c_{a} (t)$ at time $t\in [T]$:
\begin{equation}
 \label{eq: regularized quadratic loss}
\begin{aligned}
               \bm c_{a} (t) 
                =
                \argmin \limits_{\bm c_{a} \in \mathbb{R}^{d_a}}
                \sum_{(u, a, r) \in \mathcal{H}_t}
                \left[\bm c_{a}^{\intercal}  \bm \Theta_p \bm c_{u} - r(u,a,p) \right]^2 + \lambda \| \bm c_{a} \|_2^2,
\end{aligned}
\end{equation}
where $\hat{r}_{u, a, p} := \bm c_{a}^{\intercal}  \bm \Theta_p \bm c_{u}$ denotes the \emph{estimated time-varying reward}, $\mathcal{H}_t := \left\{ (u_k, a_{I_k}, r_k) \right\}_{k \in [t]}$ represents the \emph{interaction history} up to time $t$, $(u_k, a_{I_k}, r_k)$ denotes that the policy recommended item $a_{I_k}$ to user $u_k$ at time $k$ and received a reward $r_k$, and $\lambda >0$ is the regularization parameter.

To reduce the uncertainty of user preference estimations, we introduce the observed item features. Specifically, we split the context $\bm c_a (t)$ at time $t$ of item $a \in \mathcal{A}_t$ into two parts, represented as $\bm c_a (t):= [\bm s_a^\intercal, \bm x_a (t)^\intercal]^\intercal \in \mathbb{R}^{d_a}$, which includes: the observed features $\mathbf{s}_a \in \mathbb{R}^{o_a}$, and the latent features $\mathbf{x}_a(t) \in \mathbb{R}^{l_a}$ that needs to be learned online, where $d_a = o_a + l_a$. Accordingly, we redefine the estimated user preference matrix as $ \bm \Theta_p= \left[\bm \Theta_p^{s \intercal}, \bm \Theta_p^{x \intercal} \right]^\intercal$, where $\bm \Theta_p^{s} \in \mathbb{R}^{o_a\times d_u}$ corresponds to the observed item features $\bm s_a$, and $\bm \Theta_p^{x} \in \mathbb{R}^{l_a\times d_u}$ corresponds to the latent item features $\bm x_a(t)$.
As a result, we can rewrite the ridge regression  in Eq.~\eqref{eq: regularized quadratic loss} as follows:
\begin{align}
              &\hphantom{{}={}}
               \bm x_{a} (t) 
               \label{eq: regularized quadratic loss enhanced}
               \\
                &=
                \argmin \limits_{\bm x_{a} \in \mathbb{R}^{l_a}}
                \sum_{(u, a, r) \in \mathcal{H}_t}
                \left\{[\bm s_a^\intercal, \bm x_a(t)^\intercal]  \bm \Theta_p \bm c_{u} - r(u,a,p) \right\}^2 + \lambda \| \bm x_{a} \|_2^2
                \nonumber
                \\
                &=
                \argmin \limits_{\bm x_{a} \in \mathbb{R}^{l_a}}
                \sum_{(u, a, r) \in \mathcal{H}_t}
                \left[\bm s_a^\intercal \bm \Theta_p^s\bm{c}_u +\bm x_a(t)^\intercal \bm \Theta_p^x \bm{c}_u - r(u,a,p) \right]^2 +  \lambda \| \bm x_{a} \|_2^2.
                \nonumber
\end{align}

To solve the ridge regression Eq.~\eqref{eq: regularized quadratic loss enhanced}, we can easily derive the closed-form solutions as $\bm x_{a,t} = \left( \bm \Psi_{a,t} \right)^{-1} ~\bm b_{a,t}$, where 
\begin{equation*}
\begin{aligned}
    \bm \Psi_{a, t} &= \sum_{u \in \mathcal{U}_{a, t}} \left(\bm \Theta_{p}^x \bm c_u\right) \left(\bm \Theta_{p}^x \bm c_u\right)^\intercal + \lambda \bm I_{l_a}, 
   \\
    \bm b_{a,t} &= \sum_{(u, a, r) \in \mathcal{H}_t} \bm \left(\bm \Theta_{p}^x \bm c_u\right)\left[  r(u, a, p) - \left(\bm \Theta_{p}^s \bm c_u \right)^\intercal \bm s_{a}\right],
\end{aligned}
\end{equation*}
where $\mathcal{U}_{a, t}$ denotes the set of users (possibly with duplicates) who have been recommended item $a$ until time $t$, and $\bm I_{l_a} \in \mathbb{R}^{l_a \times l_a}$ is a identity matrix. The statistics $\left(\bm \Psi_{a,t}, \bm b_{a,t}\right)$ can be updated incrementally and the detailed computation can be found in Algorithm~\ref{alg:hyperbandit}.




According to the UCB policy in bandit algorithms \cite{li2010contextual, wang2017factorization,Zhang2021Counterfactual}, we define the following UCB-based relevance score function for executing action (i.e., online recommendation) at time $t$:  
\begin{equation*}
\begin{aligned}
f_t (\bm c_u, \bm c_a(t)):= \left[ \bm s_{a}^{\intercal}, \bm x_{a}(t)^{\intercal} \right] \bm \Theta_{p} \bm c_{u} 
            +
            \alpha \left[\left(\bm \Theta_{p}^{x} \bm c_{u} \right)^\intercal \left(\bm \Psi_{a, t} \right)^{-1} \bm \Theta_{p}^{x} \bm c_{u} \right]^{\frac{1}{2}},   
\end{aligned}
\end{equation*}
where $\alpha > 0$ is the exploration parameter, and the term multiplied by $\alpha$ is the exploration term. 
In this way, the executed action at time $t$ can be selected by $a_{I_t} = \argmax_{a \in \mathcal{A}_t} f_t (\bm c_u, \bm c_a (t))$.
\begin{algorithm}[!htb]
    \caption{HyperBandit}   
    \label{alg:hyperbandit}
    \begin{algorithmic}[1]
    \REQUIRE   Latent features of items $\bm x_{a\in \mathcal{A}} = \bm{0}^{l_a}$,data buffer $\mathcal{D}_{n=1}= \emptyset$,
    $\bm \Phi_{a \in \mathcal{A}, t=1} = \bm{O}^{l_a\times l_a}$, $\bm b_{a \in \mathcal{A}, t=1} =\bm{0}^{l_a}$, $\{T_n\}_{n \in [N]}$set of time steps in each updating part, regularization parameter $\lambda > 0$, exploration parameter $\alpha > 0$.  
    \STATE Initialize hypernetwork parameters $\boldsymbol{\xi}_{n=1}$ with Xavier Normal
    \FOR{$n \in [N]$}
    \FOR{$t = 1 $ to $ T_n $}
            \STATE Receive the user $u_t$ and the time period embedding $\bm s_{p_t}$
            \STATE Obtain the set of candidate items $\mathcal{A}_t$
            \STATE Obtain the observed features $\bm s_a, \forall a \in \mathcal{A}_{t}$
            \STATE Obtain the latent features $\bm x_a(t), \forall a \in \mathcal{A}_{t}$
            \STATE Estimated the user preference matrix $\bm \Theta_{p_t}^{(n)} :=\left[ \bm \Theta_{p_t}^{s(n)\intercal}, \bm \Theta_{p_t}^{x(n)\intercal} \right]^\intercal \leftarrow h_{\bm \xi_{n}} (\bm s_{p_t})$
            \STATE Recommend item $a_{I_t} \in \mathcal{A}_{t}$ to user $u_t$ following $a_{I_t} \leftarrow \arg \max _{a \in  \mathcal{A}_t}
            \left[ \bm s_{a}^{\intercal}, \bm x_{a}(t)^{\intercal} \right] \bm \Theta_{p_t}^{(n)} \bm c_{u_t} 
            +
            \alpha \left[\left( \bm \Theta_{p_t}^{x(n)} \bm c_{u_t} \right)^\intercal \left(\bm \Psi_{a, t} \right)^{-1} \bm \Theta_{p_t}^{x(n)} \bm c_{u_t} \right]^{\frac{1}{2}}$ 
            
            \STATE Observe reward $r_t = r(u_t,a_{I_t},p_t)$
            \STATE $\mathcal{D}_{n} \leftarrow \mathcal{D}_{n} \cup \left\{ (u_t, a_{I_t}, p_t, r_t,\mathcal{A}_t)  \right\} $
            \STATE  $// ~~ \texttt{Bandit Policy Updating}$            
            \STATE Get the user preference vector $\bm P_{t} \leftarrow \left(\bm \Theta_{p_t}^{x(n)} \bm c_{u_t}\right)^\intercal  \in \mathbb{R}^{l_a}~$ for the latent item features

            \STATE Get the user preference vector $\bm Q_{t} \leftarrow \left(\bm \Theta_{p_t}^{s(n)} \bm c_{u_t}\right)^\intercal \in \mathbb{R}^{o_a}$ for the observed item features
     
            \STATE $\bm \Phi_{a_{I_t},t+1} \leftarrow \bm \Phi_{a_{I_t},t} + \bm P_{t}^\intercal \bm P_{t} $, \quad  $\bm \Psi_{a_{I_t},t+1} \leftarrow \lambda \bm I + \bm \Phi_{a_{I_t},t+1}$
            
            \STATE $\bm b_{a_{I_t},t+1} \leftarrow \bm b_{a_{I_t},t} + \bm P_{t}^\intercal 
            \left(
                \bm r_t - \bm Q_{t} \bm s_{a_{I_t}}
            \right)$
            
            \STATE $\bm x_{a_{I_t},t+1} \leftarrow 
                \left( \bm \Psi_{a_{I_t},t+1}  \right)^{-1} ~\bm b_{a_{I_t},t+1}$
            \ENDFOR
            
            \STATE  $// ~~ \texttt{     Hypernetwork Updating}$
        
            \STATE Update hypernetwork parameter $\bm \xi_{n+1} \leftarrow \Delta (\bm \xi_{n})$ using efficient training method via low-rank factorization (in Sec.~\ref{sec:HB:Lowrank:training}) and the Adam optimizer on $\mathcal{D}_{n}$
            \STATE Release $\mathcal{D}_{n}$
            and set $\mathcal{D}_{n+1}\leftarrow \emptyset $
    \ENDFOR
\end{algorithmic}
\end{algorithm}

\subsubsection{Hypernetwork for Time-Varying Preference}
\label{sec:HB:Hypernetwork}
In the last section, we describe the bandit policy given parameter matrix $\bm \Theta_p$ in time period $p$. In this section, we explain how the hypernetwork generates the parameter matrix $\bm \Theta_p$. 
The main concept involves utilizing a hypernetwork that takes the embedding of the current time period as input and generates the parameters of the user preference matrix in the bandit policy. This enables the policy to adapt and adjust itself to accommodate changes in the distribution of user preferences over time.


To ensure stability in online recommendation, we incrementally update the hypernetwork $h$ in mini-batches, where the total $T$ time steps are divided into $N$ parts, and the $n$-th part, $n \in [N]$, contains $T_n$ time steps, corresponding to $T_n$ interaction histories.
In this way, the hypernetwork $h$ is updated $N$ times, and during the $n$-th update, the data buffer $\mathcal{D}_{n}:= \left\{ (u_t, a_{I_t},p_t, r_t, \mathcal{A}_t) \right\}_{t \in [T_n]}$ is used as the training data\footnote{It is important to note that the interaction history corresponding to the same index $t$ in different data buffers $\mathcal{D}_n$ may be different.}, where $r_t = r(u_t,a_{I_t},p_t)$. Then, given the time period embedding $\bm s_p$, the hypernetwork after the $(n-1)$-th update, denoted by $h_{\bm \xi_n}$, can be represented by:
\begin{equation}
\label{eq:hn:formulation}
\bm \Theta_{p}^{(n)}  := 
h_{\bm \xi_{n}} (\bm s_p),
\end{equation}
where $\bm \xi_n$ represents the model parameters of the hypernetwork, and the superscript $(n)$ on $\bm \Theta_{p}^{(n)}$ indicates that it is generated by $h_{\bm \xi_n}$. 
As illustrated in Figure~\ref{fig:hyperbandit_structure}, we implement the hypernetwork $h$ using a Multi-Layer Perceptron (MLP). 
In this way, the MLP acts like a condition network, inputting the embedding of current time period, outputting the corresponding user preference matrix. 
Besides, the time period embedding $\bm s_p$ is generated via GloVe model~\cite{pennington2014glove} by imputing the current time period $p$.



To train the hypernetwork, we take inspiration from the Listnet loss design~\cite{cao2007learning} to quantify the discrepancy between the estimated reward and the true label for each candidate item in $\mathcal{A}_t$ at time~$t$. 
Specially, we use the estimated time-varying reward $\hat{r}_{u, a, p} = \bm c_{a}^{\intercal}  \bm \Theta_p \bm c_{u}$ in Eq.~\eqref{eq: regularized quadratic loss}, 
and construct the true labels according to the following rules: if the user $u$ clicks on the recommended item $a$ in time period $p$, the label is set to 1; if the user skips the recommended item, the label is set to -1; if the item is a candidate but not recommended, the label is set to 0. Formally, $y_{u, a, p} = 1$ if $r_{u, a, p} = 1$; $y_{u, a, p} = -1$ if $r_{u, a, p}$ = 0; $y_{u, a, p} = 0$ if it is a candidate item but not recommended. Then, assuming that the number of actions is $M := |\mathcal{A}_1| = \cdots = |\mathcal{A}_T|$, 
during the $n$-th incremental update, the loss function on the data buffer $\mathcal{D}_{n}$ could be shown as follows: 
\begin{equation*}  
\mathcal{L}_{\bm \xi}^{(n)} = -
\sum_{t=1}^{T_n} 
\sum_{k=1}^{M} P \left(y_{u_t, a_k, p_t}\right) \log \widehat{P} \left(\hat{r}_{u_t, a_k, p_t}\right),
\end{equation*}
where $\bm \xi$ represents the hypernetwork parameters that need to be optimized, $p_t$ corresponds to the time period where index $t$ in $\mathcal{D}_n$ is located, and
\begin{equation*}    \widehat{P} \left(\hat{r}_{u_t, a_k, p_t}\right)=\frac{\exp \left(\hat{r}_{u_t, a_k, p_t}\right)}{\sum_{i=1}^{M} \exp \left(\hat{r}_{u_t, a_i, p_t}\right)}, ~     
{P} \left(y_{u_t, a_k, p_t}\right)=\frac{\exp \left(y_{u_t, a_k, p_t}\right)}{\sum_{i=1}^{M} \exp \left(y_{u_t, a_i, p_t}\right)}.
\end{equation*}


\subsection{\mbox{Efficient Training via Low-Rank Factorization}}
\label{sec:efficienttraining}

\subsubsection{Analysis of Low-Rank Structure of User Preference Matrix}

\label{subsec: Low-rank Factorization}

\begin{figure*}[!thp]
\centering
\begin{subfigure}[b]{0.32\textwidth}
\includegraphics[width=\textwidth]{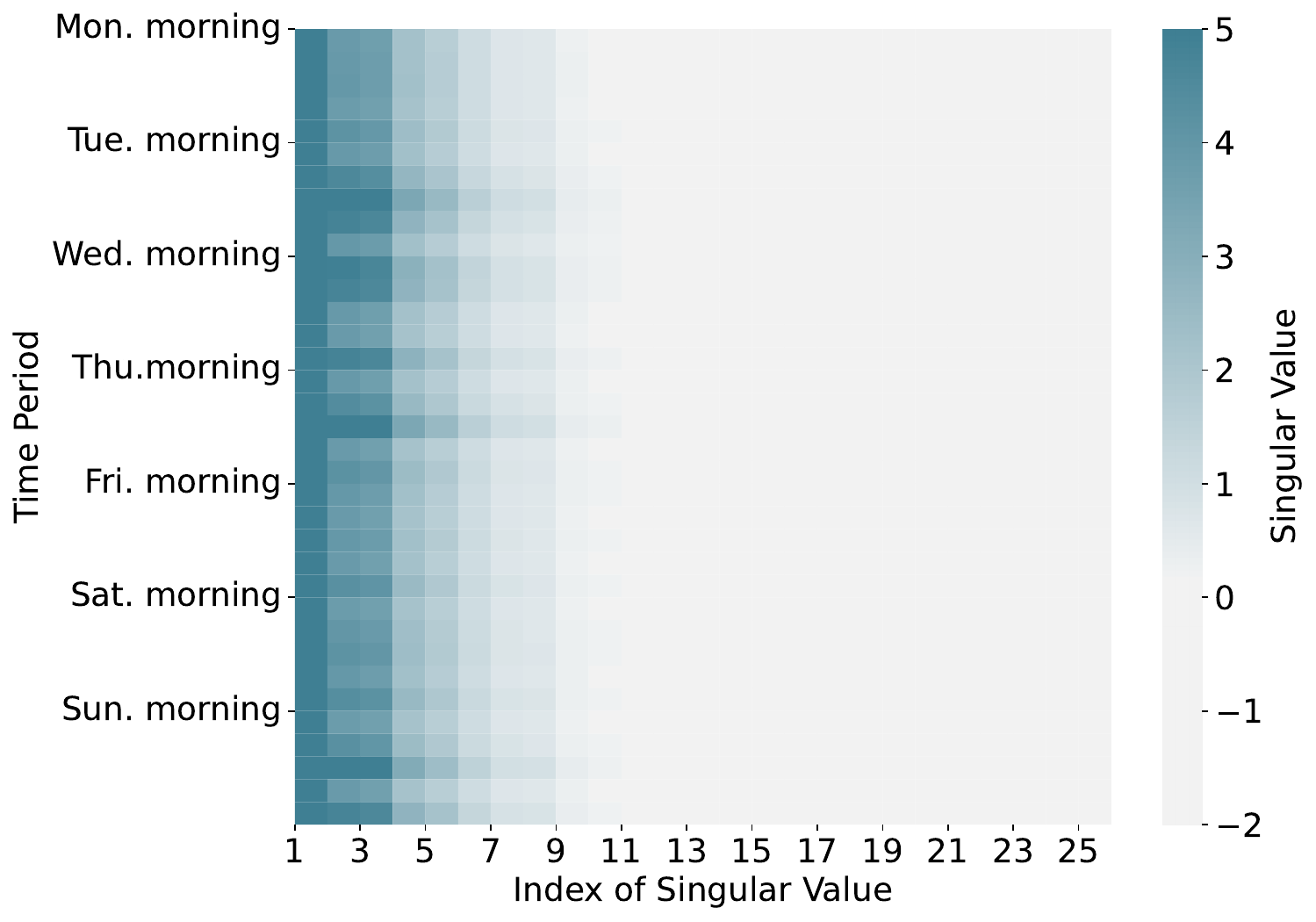}
\caption{KuaiRec}
\end{subfigure}
\hfill
\begin{subfigure}[b]{0.32\textwidth}
\includegraphics[width=\textwidth]{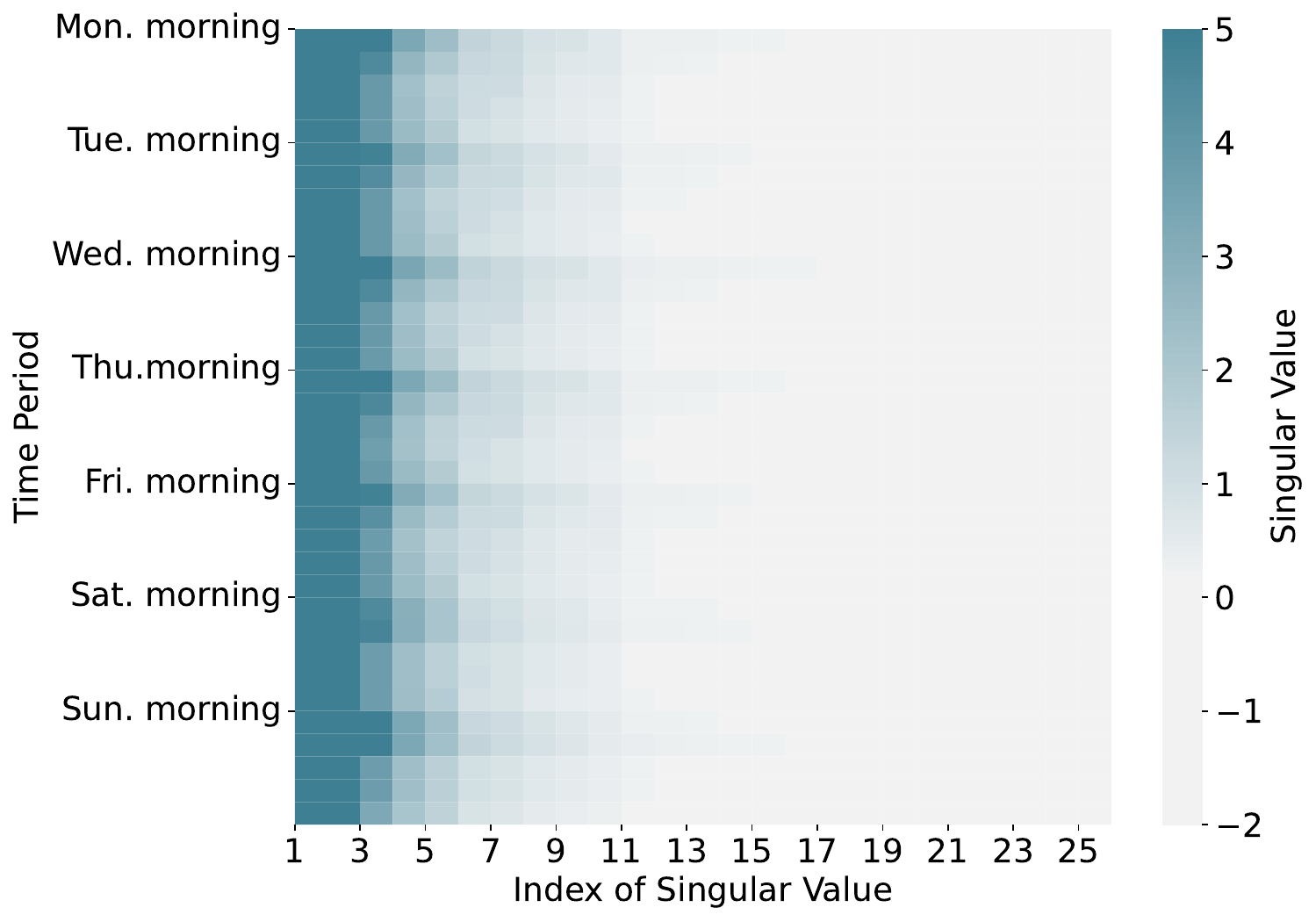}
\caption{NYC}
\end{subfigure}
\hfill
\begin{subfigure}[b]{0.32\textwidth}
\includegraphics[width=\textwidth]{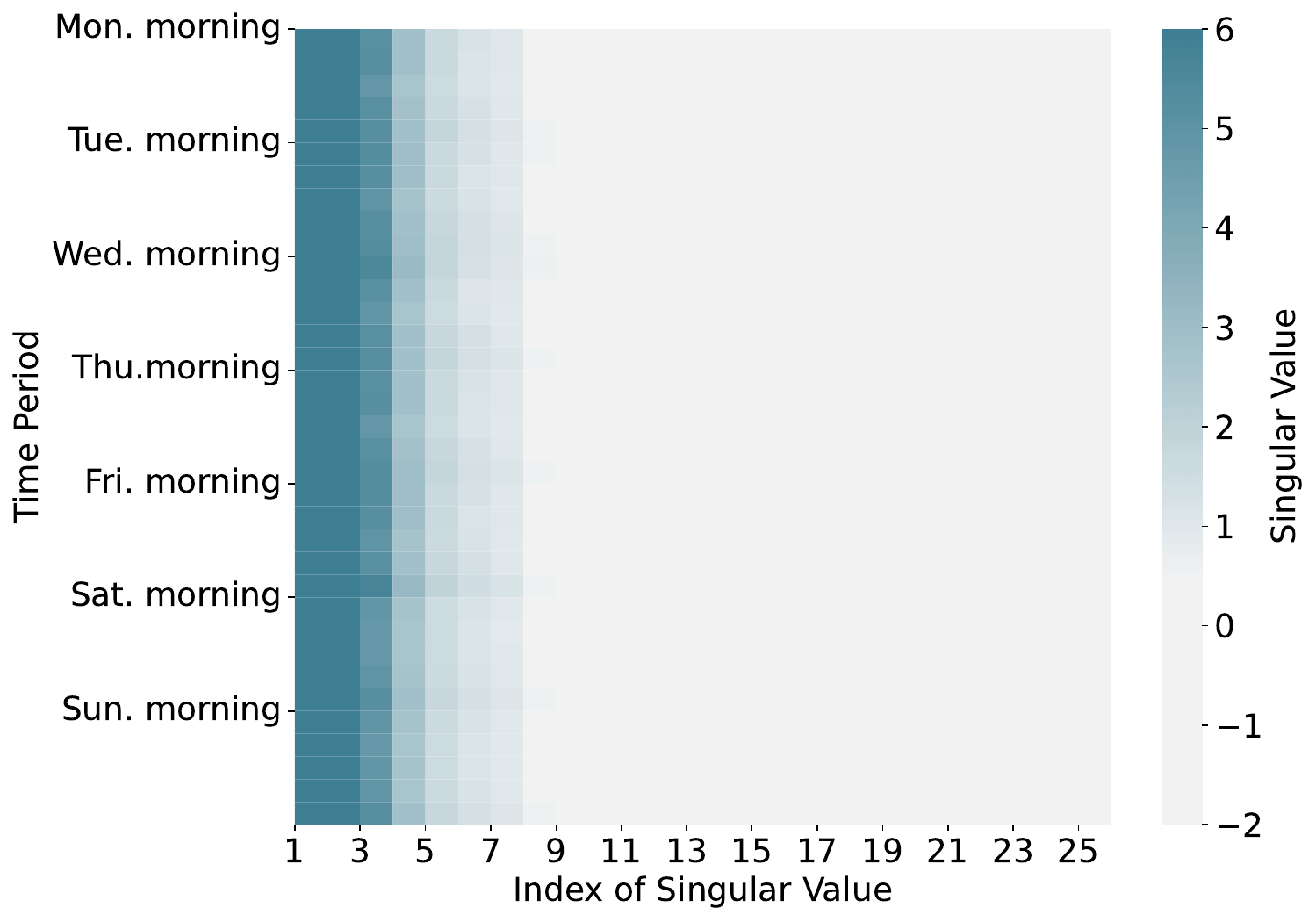}
\caption{TKY}
\end{subfigure}
\vspace{-1ex}
\caption{
The distribution of singular eigenvalues (SEs) of user preference matrices across different time periods. The horizontal axis represents the index of SEs, arranged in descending order, while the vertical axis represents the time periods. The darkness of the colors corresponds to the magnitude of the singular values.}
\label{fig:SVD}
\end{figure*}
Since the user preference matrix $\bm \Theta_{p} \in \mathbb{R}^{d_a \times d_u}$ is generated by the hypernetwork in Eq.~\eqref{eq:hn:formulation}, a large output dimension (i.e., $d_a \times d_u$) 
would incur significant training costs. Hence, we consider representing the entire user preference matrix using a smaller number of parameters. 
Based on this motivation, it is natural to investigate whether the user preference matrix $\bm \Theta_p$ exhibits a low-rank structure. 
To verify the presence of low-rank structures, we perform singular value decomposition (SVD) on $\bm \Theta_{p}^{(N)}$ across different time periods. 
As shown in Fig.~\ref{fig:SVD}, when the singular values of the user preference matrices are sorted in descending order for different time periods, it becomes apparent that the distribution of singular values is concentrated in the first few dimensions, which is always less than half of the dimensionality of the user preference matrices\footnote{Following the setting of baselines~\cite{li2010contextual,wu2018learning,hariri2015adapting,wang2017factorization}, we set $d_a = d_u = 25$, thus $\bm\Theta_p$ is a square matrix. 
}. 
Based on this observation, we can conclude that there are strong low-rank structures present in the user preference matrices. This implies that a low-rank representation of the matrix $\bm \Theta_p$ could preserve nearly all of its information content.

\subsubsection{Training Process with Low-Rank Factorization}
\label{sec:HB:Lowrank:training}
\begin{figure}[!bhtp]
    \centering
    \includegraphics[width=\linewidth]{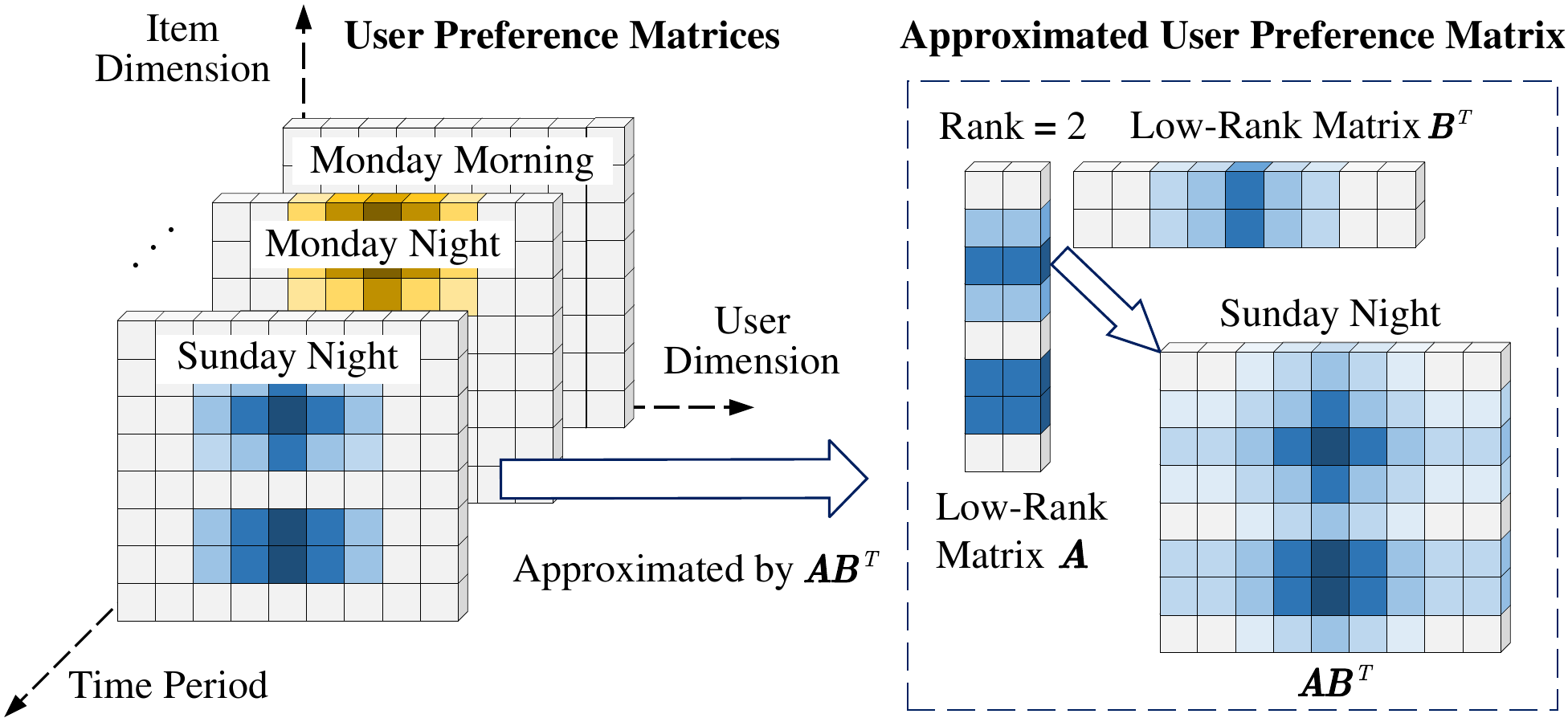}
    \caption{User preference matrix estimation using low rank factorization: An example with estimated rank $\tau = 2$.}
    \label{fig:lowrank_structure}
\end{figure}

Based on the analysis above, we try to improve the training efficiency of hypernetwork through explicitly modeling the low-rank structure in the user preference matrix $\bm \Theta_p$ (for ease of exposition, we omit the superscript of $\bm \Theta_p^{(n)}$ below). Specifically, we propose to approximate  $\bm\Theta_p$ with its low-rank approximation. Here, we leverage matrix factorization approach to achieve the approximation. 
Given the \emph{estimated rank} $\tau>0$, we model the low-rank structure of $\bm \Theta_p$ with the product of two rank-$\tau$ latent matrices $\bm A_p \in \mathbb{R}^{d_a\times \tau}$ and $\bm B_p\in \mathbb{R}^{d_u\times \tau}$, i.e., $\bm \Theta_p \approx \bm A_p  \bm B_p^\intercal$, as shown in Fig.~\ref{fig:lowrank_structure}. 



In the implementation of the hypernetwork, given a time period $p \in \mathcal{P}$, the hypernetwork outputs a vector represented by $\mathrm{Concat}(\mathrm{Vec}(\bm A_p),\mathrm{Vec}(\bm B_p) ) \in \mathbb{R}^{\tau d_a+\tau d_u}$, where $\mathrm{Concat}(\cdot)$ denotes the concatenation operation. Then, the vector $\mathrm{Vec}(\bm A_p) \in \mathbb{R}^{\tau d_a}$ is reshaped into a matrix  $\bm A_p \in \mathbb{R}^{d_a \times \tau}$, and the vector $\mathrm{Vec}(\bm B_p) \in \mathbb{R}^{\tau d_u}$ is reshaped into a matrix $\bm B_p \in \mathbb{R}^{d_u \times \tau}$. Finally, the product $\bm A_p  \bm B_p^\intercal$ is obtained to estimate $\bm \Theta_p$.  This matrix factorization reduces the output dimension of the hypernetwork $h$ (defined in Eq.~\eqref{eq:hn:formulation}) from $d_a  d_u$ to $\tau  (d_a + d_u)$, 
effectively alleviating the training efficiency issues.

\section{Regret Analysis}
The regret bound serves as a fundamental theoretical guarantee for online learning algorithms \cite{Cesa2006Prediction,Bubeck2012Regret,Shalev2011OLA,Hazan2016Introduction,Zhang2019Survey}. 
In this section, we provide a regret bound of the proposed HyperBandit. 
First, we define the \emph{regret} as follows:
\begin{equation}
\label{eq:HB:regret:def}
    \mathrm{Reg}(T) := 
    \sum_{t \in [T]} 
    \left[
        r^*(u_t, a_t^*, p_t) - r^*(u_t, a_{I_t}, p_t) 
    \right],
\end{equation}
where $a_t^*$ represents the action with the highest time-varying true reward $r^*$ (defined in Eq.~\eqref{eq:hb:time_reward}) at time $t$, $u_t$ denotes the user for whom the item is recommended at time $t$, and $p_t$ represents the time period to which $t$ belongs.  
Recalling that $I_t$ denotes the index of the action executed by HyperBandit at time $t$, the regret in Eq.~\eqref{eq:HB:regret:def} measures the difference between the accumulated time-varying true rewards of the best policy and our policy.

\begin{figure*}[!thbp]
\centering
\begin{subfigure}[b]{0.32\textwidth}
\includegraphics[width=\textwidth]{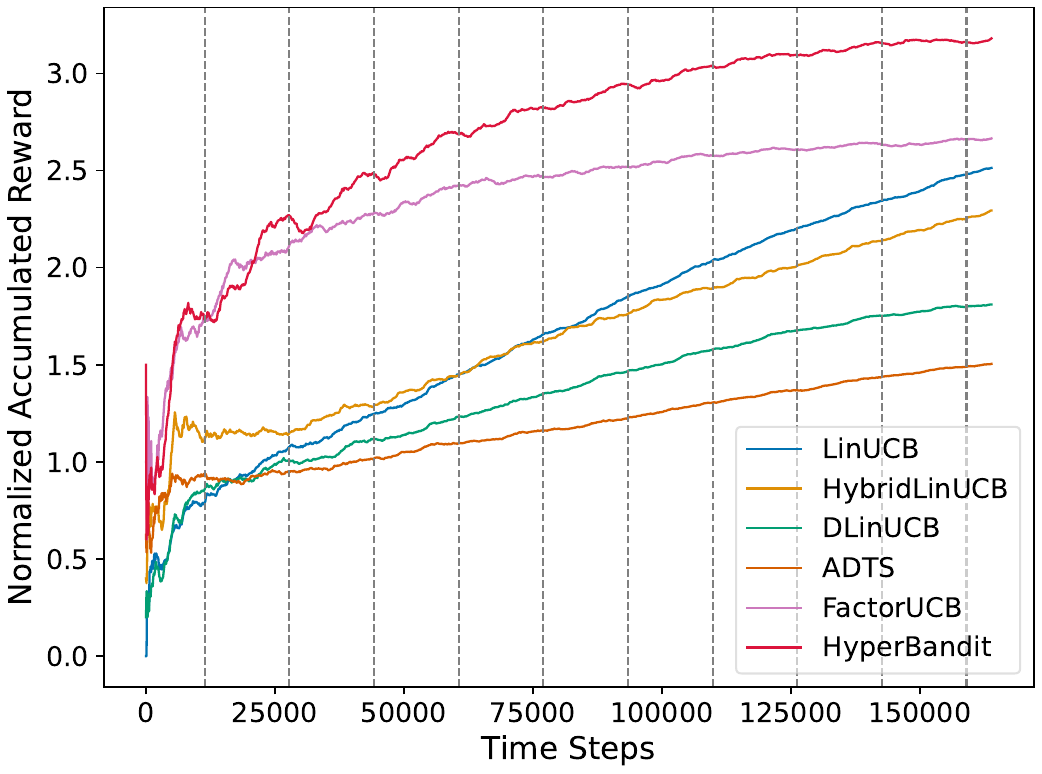}
\caption{ KuaiRec}
\label{subfig:reward:a}
\end{subfigure}
\hfill
\begin{subfigure}[b]{0.32\textwidth}
\includegraphics[width=\textwidth]{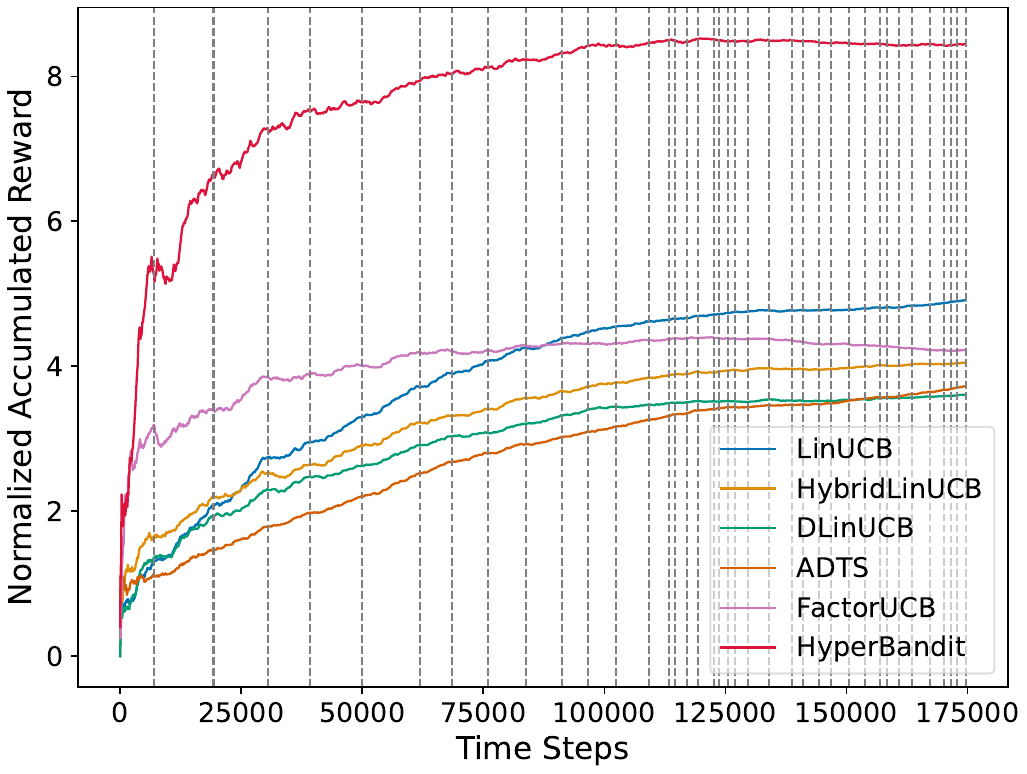}
\caption{ NYC}
\label{subfig:reward:b}
\end{subfigure}
\hfill
\begin{subfigure}[b]{0.32\textwidth}
\includegraphics[width=\textwidth]{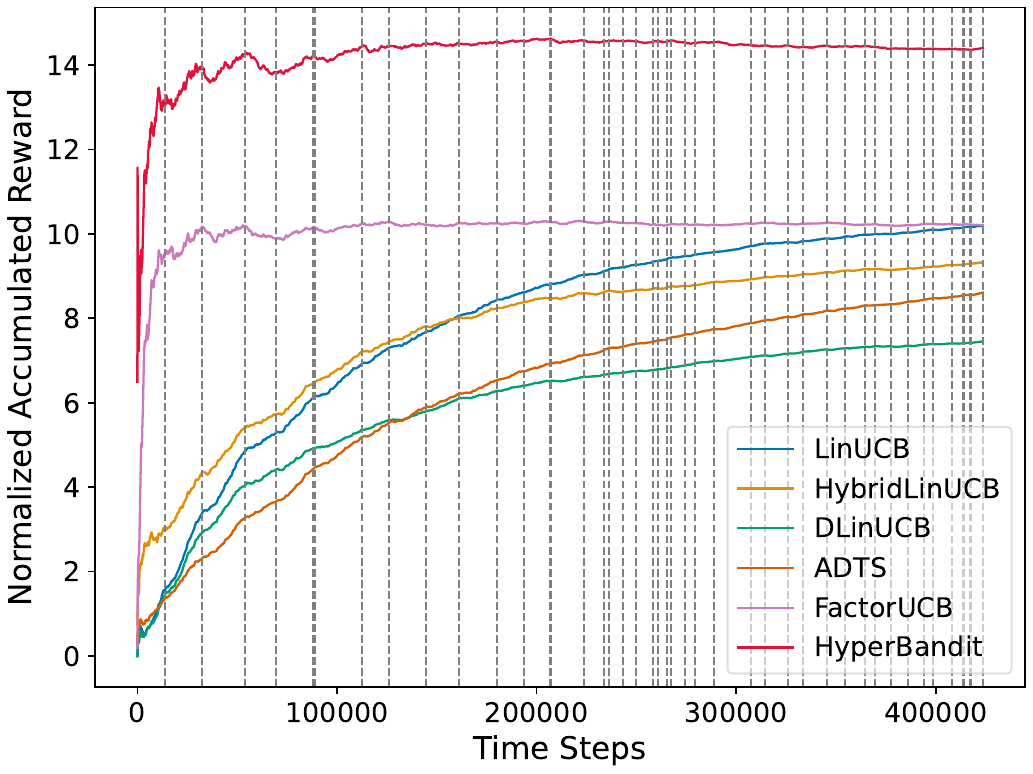}
\caption{ TKY}
\label{subfig:reward:c}
\end{subfigure}
\vspace{-3ex}
\caption {Normalized accumulated reward of baselines, and the proposed HyperBandit on three datasets, KuaiRec \& NYC \& TKY. Note that The grey dashed lines represent the boundaries between weekdays and weekends. The $x$-axis represents the interaction data arranged in chronological order, and the $y$-axis represents the normalized accumulated reward. }

\label{fig:reward}
\end{figure*}

\begin{theorem}[Regret Upper Bound of HyperBandit]
\label{thm:hyperbandit:regret}
Assume that the dimension of the latent features is $l_{a} = L, \forall a \in \mathcal{A}.$
The sequence of the actions executed by HyperBandit enjoys the following regret upper bound: with probability at least $1 - \delta$, 
\begin{equation}
\label{eq:HB:regret:bound}
    \mathrm{Reg}(T) \leq
    2 C_{\bm x} \sqrt{2 L T 
    \ln\left(1 + \frac{C_{\bm \Theta} C_{\mathcal{U}}^2 T}{2\lambda L \delta}\right)} +
    C_{\bm x} \sqrt{\frac{2}{\lambda}}
    \sum_{n \in [N] } E_n,
\end{equation}
where 1).~
$C_{\bm x} = \max_{a \in \mathcal{A}_T} \|  \bm x_a (T) - \bm x_a^*\|_{\bm \Psi_{a, T}}$,
$\|\bm x\|_{\bm \Psi} := \sqrt{\bm x^\intercal \bm \Psi \bm x}$ denotes the elliptic norm of $\bm x$ with respect to the matrix $\bm \Psi$,
$\bm x_a^*$ is the true latent features of item $a$, 
and $C_{\bm \Theta} = \max_{n \in [N], p \in \mathcal{P}} \left\|\bm \Theta_p^{(n)} \right\|_{\mathrm{F}}^2$, $C_{\mathcal{U}} = \max_{u \in \mathcal{U}} \|\bm c_u\|_2^2$;
2).~ $E_n := \sum_{i\in [T_n]}\left\|\left(\bm \Theta_{p_{i, n}}^{(n)} - \bm \Theta_{p_{i, n}}^*\right) \bm c_{u_{i, n}} \right\|_2$ denotes the error caused by the hypernetwork updated $N$ times using $T_n$ examples in the $n$-th update, and the subscript $(\cdot)_{i, n}$ denotes the time step $i \in [T_n]$ after the $(n-1)$-th update.
\end{theorem} 
    The error $E_n$  in  Eq.~\eqref{eq:HB:regret:bound} caused by the hypernetwork can be decomposed into the sum of the following three parts. 
    1). {\bf Approximation error}  measures the discrepancy between the optimal hypothesis in hypernetwork space and the target function that generates $\bm \Theta_p^* \bm c_u $ in $E_n$.
    From the results in    \cite{galanti2020modularity}, 
    we obtain that the approximation error of the hypernetwork $h$ (defined in Eq.~\eqref{eq:hn:formulation}) with ReLU activation function is $\varepsilon$, providing that the number of trainable parameters in the hypernetwork is $\Omega \left( \varepsilon^{- U/S^*} + \varepsilon^{- P/S^*} \right)$, 
    where we assume $d_{u} = U, \forall u \in \mathcal{U},$ $d_{p} = P, \forall p \in \mathcal{P}$, and $S^*$ denotes the order of smoothness of the target function. 
    2).~{\bf Optimization error}: measures the accumulated deviation between the hypernetwork parameters obtained through the online optimization algorithm and those of the optimal hypothesis in the hypernetwork space. Our HyperBandit equipped with a mini-batch first-order optimization method incurs an accumulated optimization error of order $O(T/N)$, assuming that each data buffer $\mathcal{D}_n, n \in [N]$ contains an equal number of examples. 
    3).~{\bf Estimation error} measures the  the error caused by the estimated user preference matrix using low-rank factorization. 
    According to the analyses in Sec.~\ref{subsec: Low-rank Factorization}, the user preference matrix exhibits a low-rank structure. Assuming that the maximum rank of the user preference matrices is $R$, if the estimated rank $\tau$ in the low-rank factorization discussed in Sec.~\ref{sec:HB:Lowrank:training} is set to $\tau \geq R$, and the best rank-$\tau$ approximation can be obtained, then the estimation error would be zero. 

    Setting the number of trainable parameters in the hypernetwork as $\Omega \left(\sqrt{T}^{U/S^*} + \sqrt{T}^{ P/S^*}\right)$, the number of hypernetwork training iterations $N = O(\sqrt{T})$, and the estimated rank $\tau \geq R$, we can derive an upper bound for the error term $\sum_{n \in [N] } E_n$ in Eq.~\eqref{eq:HB:regret:bound} of order $\widetilde{O}(\sqrt{T})$. This, in turn, leads to a sublinear regret upper bound of order $\widetilde{O}(\sqrt{T})$ for HyperBandit.






\section{Experiments}\label{sec:Experiments}
We conducted experiments to evaluate the performance of HyperBandit on datasets for short video recommendation and point-of-interest (POI) recommendation.

\begin{table*}[!th]
    \centering
    \caption{Comparisons of normalized accumulated reward, running time (sec., mean) and training time (sec., mean) of hypernetwork on KuaiRec, Foursquare (NYC) and Foursquare (TKY). The ``Running Time of BP'' means the average time cost of online recommendation and updating by Bandit Policy at each time step, and the ``Training Time of HN'' means the average time cost for training HyperNetwork at each time step. ``--'' means the corresponding algorithm has no hypernetwork. }
    \vspace{-1ex}
    \label{tab:comparisons of NAR and RT}
    \begin{subtable}[ht]{1\linewidth}
    \centering
   \begin{spacing}{1}
    \begin{tabular}{l|c|c|c|c|c|c|c}
        \hline
        \multirow{2}{*}{Algorithm} & \multicolumn{3}{c|}{Normalized Accumulated Reward} & \multicolumn{2}{c|}{Running Time of BP }&\multicolumn{2}{c}{Training Time of HN } \\
        \cline{2-8}
                                   & KuaiRec & NYC & TKY  & KuaiRec & Foursquare & KuaiRec & Foursquare \\ 
        \hline \hline
        LinUCB &$2.56\pm 0.04$ &$4.86\pm0.05$ & $10.15\pm0.10$& $3.09\mathrm{e-}04$ & $3.08\mathrm{e-}04$& -- & -- \\
        HybridLinUCB & $2.29\pm0.03$&$4.05\pm0.07$ &$9.33\pm0.03$ &$2.52\mathrm{e-}02$& $2.65\mathrm{e-}02$& -- & -- \\
        DLinUCB &$1.84\pm0.03$ &$3.63\pm0.08$&$7.39\pm0.08$ & $3.51\mathrm{e-}04$ & $3.55\mathrm{e-}04$& -- & -- \\
        ADTS &$1.50\pm0.02$ &$3.80\pm0.10$ &$8.63\pm0.09$ & $6.52\mathrm{e-}03$ & $6.69\mathrm{e-}03$& -- & -- \\
        FactorUCB &$2.70\pm0.05$ &$4.19\pm0.04$ &$10.22\pm0.03$ & $1.01\mathrm{e-}01$ & $1.11\mathrm{e-}01$& -- & -- \\
        \hline
        \textbf{HyperBandit} ($\tau=1$) &$\mathbf{3.79\pm0.18}$ &$6.46\pm0.45$ &$13.37\pm 0.22$ & $1.65\mathrm{e-}03$& $1.62\mathrm{e-}03$&$1.45\mathrm{e-}04 $&$ 1.06\mathrm{e-}04$\\  
        \textbf{HyperBandit} ($\tau=5$) &$3.51\pm0.06$ &$8.08\pm0.09$ &$13.99\pm 0.29$ & $1.60\mathrm{e-}03$& $1.63\mathrm{e-}03$&$1.58\mathrm{e-}04$ & $1.21\mathrm{e-}04$\\ 
        \textbf{HyperBandit} w/o Low-Rank &$3.24\pm0.11$ &$\mathbf{8.27\pm0.17}$ &$\mathbf{14.49\pm 0.07}$ & $1.63\mathrm{e-}03$& $1.62\mathrm{e-}03$& $1.72\mathrm{e-}04$& $1.73\mathrm{e-}04$\\
        \hline
    \end{tabular}
    \end{spacing}
    \end{subtable}
    \vspace{-1ex}
\end{table*}

\subsection{Experimental Settings}

\subsubsection{Baselines.} HyperBandit was compared with several algorithms that construsted in stationary or piecewise-stationary environment, including:







\textbf{LinUCB \cite{li2010contextual}} is a classical contextual bandit algorithm that addresses the problem of personalized recommendation.

\textbf{HybridLinUCB \cite{li2010contextual}} is a variant algorithm of LinUCB that takes into account both shared and non-shared interests among users. 

\textbf{DLinUCB \cite{wu2018learning}} is built upon a piecewise stationary environment, where each user group corresponds to a slave model. Whether to discard a slave model is based on the detection of ``badness''. 


\textbf{ADTS \cite{hariri2015adapting}} is a bandit algorithm based on Thompson sampling, which tend to discard parameters before the last change point.


\textbf{FactorUCB \cite{wang2017factorization}} leverages observed contextual features and user interdependencies to improve the convergence rate and help conquer cold-start in recommendation.

\subsubsection{Hyperparameter Settings.}
We implemented the hypernetwork $h$ in  Eq.~\eqref{eq:hn:formulation} using a MLP. The MLP consists of 1 input layer, 8 hidden layers, and 1 output layer. 
The number of nodes in the each layer is as follows: 30, 256, 512, 1024, 1024, 1024, 1024, 512, 256, $25*\tau*2$. We applied ReLU activation function after each hidden layer. We trained the hypernetwork every 2000 time steps (i.e., $T_n = 2000, n\in [N]$) on KuaiRec and NYC, while $T_n = 5000$ on TKY. 
Early stopping is applied in training process to avoid overfitting.

For the parameters in bandit policy, we set the exploration parameter $\alpha$ to 0.1 and the regularization parameter $\lambda$ to 0.1 for all the algorithms. 
The size of the candidate item set $\mathcal{A}_t, \forall t \in [T]$ at each time step was set to 25 in all algorithms. The dimensions of both the context features of users and the context features of items were set to 25. In FactorUCB and HyperBandit, the dimensions of latent features of items were set to 10, and the dimensions of observed features of items were set to 15.

\subsubsection{Evaluation Protocol.}
The accumulated reward was utilized to assess the recommendation accuracy of algorithms, which was computed as the sum of the observed reward from the beginning to the current step. The normalized accumulated reward refers to the accumulated reward normalized by the corresponding logged random strategy.


\subsection{\mbox{Experiments on Short Video Recommendation}}
We employed KuaiRec~\cite{gao2022kuairec} for evaluation, that is a real-world dataset collected from the recommendation logs of the video-sharing mobile app Kwai\footnote{https://github.com/chongminggao/KuaiRec}.
The dense interaction matrix we used contains 1411 users, 3327 items and 529 video categories (i.e., tags). Each interactive data includes user id, video id, play duration, video duration, time, date, timestamp, and watch ratio, etc. 
Following the settings in~\cite{wan2021contextual}, we used video categories (tags) as actions. In this experiment, we treated watch ratio higher than 2.0 as positive feedback. If the action (i.e., tag) got positive feedback in other time periods while not in current period, we assumed that the current user would give a negative feedback to it.


 

To fit the data into the contextual bandit setting, we pre-processed it first. Initially, we encoded all the user features provided by KuaiRec, including user activity level, number of followed users, and others. Subsequently, we applied PCA to reduce the dimensionality of the context feature vectors. We retained the first 25 principal components and applied the same procedure to the context feature vectors of items (i.e., $d_a = d_u =25$ ). For a particular time step, the video tag id having positive feedback  was picked and the remaining 24 were randomly sampled from the tags which would get negative feedback in the current time step. Besides, we extracted data for one week from August 10th, 2020, to August 16th, 2020. From each time period within that week, we randomly sampled several time steps to reconstruct the data. This sampling process was repeated 10 times to generate data for 10 weeks. 

The results are shown in Fig.~\ref{subfig:reward:a} and Table~\ref{tab:comparisons of NAR and RT}, we can clearly notice that HyperBandit outperformed all the other baselines on KuaiRec in terms of rewards. As environment is periodic, both DLinUCB and ADTS were worse than others since these two algorithms were designed for the piece-wise stationary environment (i.e., they need to abandon the knowledge acquired during past periods). As more observations recurrent, LinUCB quickly catched up, because it is better to regard periodic environment as a stationary environment rather than piecewise environment. Besides, FactorUCB leveraged observed contextual features and dependencies among users to improve the algorithm's convergence rate, leading to good performance at the beginning.

\subsection{Experiments on POI Recommendation}

Foursquare NYC \& TKY dataset~\cite{yang2014modeling} includes long-term (about 10 months) check-in data in New York city (NYC) and Tokyo (TKY) collected from Foursquare\footnote{https://foursquare.com/} from 12 April 2012 to 16 February 2013. 
Table~\ref{tab:foursquare data info} shows the statistics of two check-in datasets: NYC and TKY. Each dataset includes user id, venue id, venue category id, venue category name, latitude, longitude, and timestamp, etc.

\begin{figure*}[!th]
\centering
\begin{subfigure}[b]{0.32\textwidth}
\includegraphics[width=\textwidth]{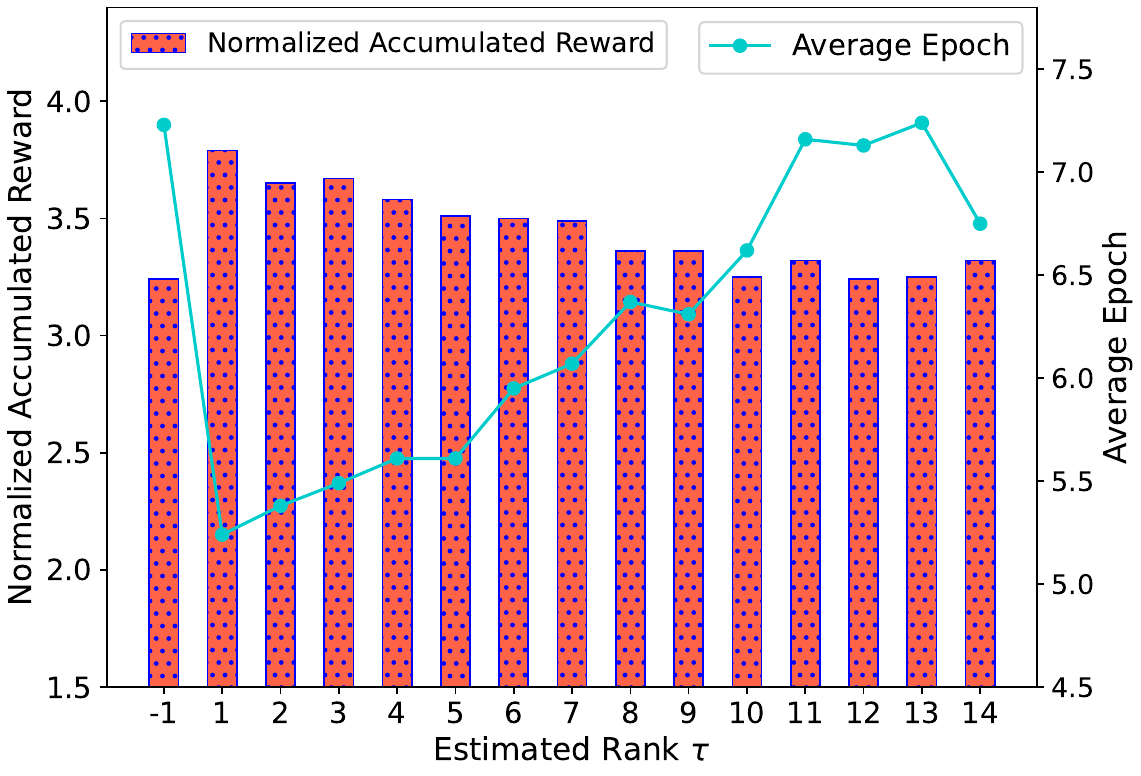}
\caption{KuaiRec}
\end{subfigure}
\hfill
\begin{subfigure}[b]{0.32\textwidth}
\includegraphics[width=\textwidth]{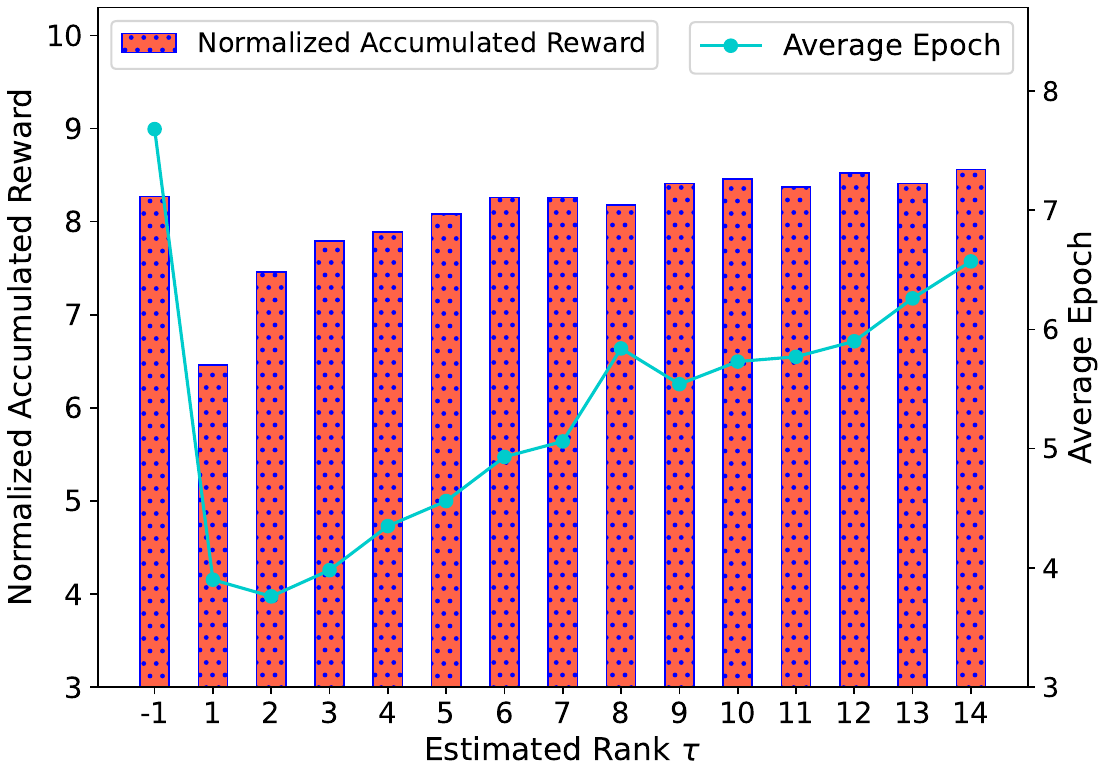}
\caption{NYC}
\end{subfigure}
\hfill
\begin{subfigure}[b]{0.32\textwidth}
\includegraphics[width=\textwidth]{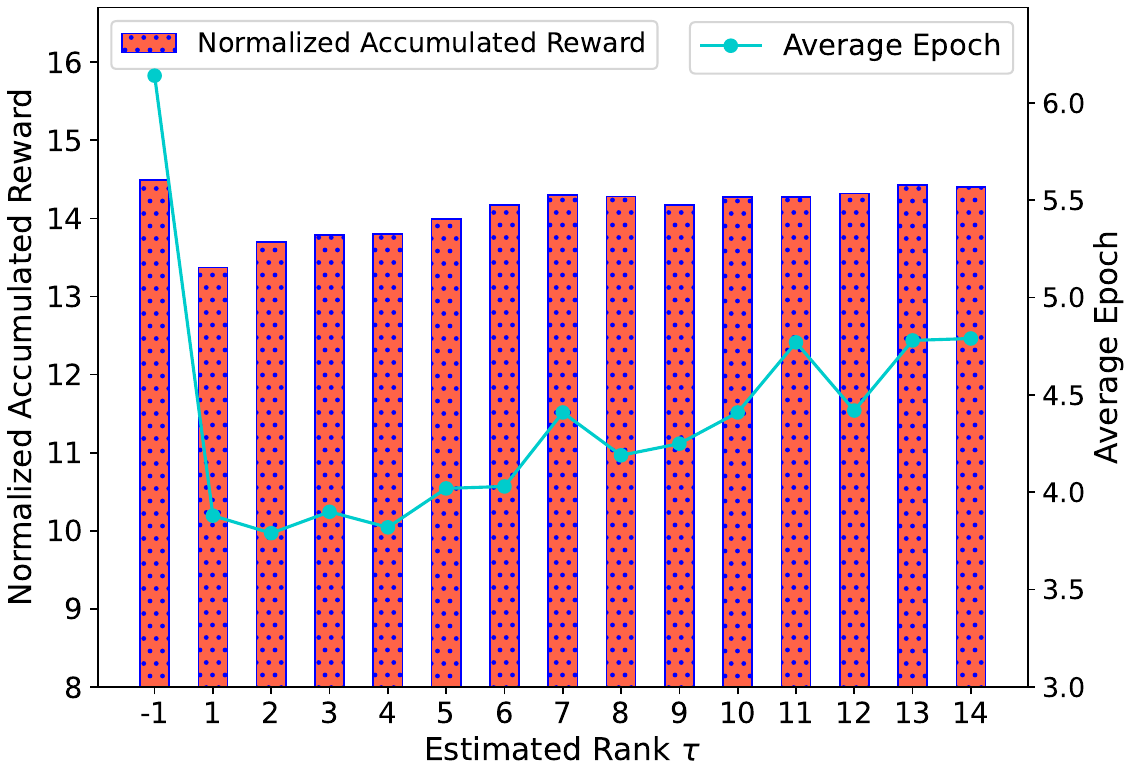}
\caption{TKY}
\end{subfigure}
\vspace{-2ex}
\caption{Performance of low-rank factorization in HyperBandit on different estimated rank across three datasets. Note that the bar chart shows normalized accumulated reward, while the line chart shows average epoch in training process (a larger average epoch indicates a longer training time of hypernetwork). The first data point (with an $x$-coordinate of ``$\mathrm{-1}$'') represents the result obtained without utilizing low-rank factorization.}
\label{fig:ranks performance}
\vspace{-2.2ex}
\end{figure*}

Similarly, we used POI categories as actions. The ground-truth categories of the check-ins were considered positive samples of the current step while the rest categories were considered negative. Initially, instead of following the setting in \cite{cesa2013gang} that used TF-IDF for representation construction, we employed the GloVe model~\cite{pennington2014glove} to craft a 300-dimensional feature vector, enhancing item context representation. Subsequently, PCA reduced vector dimensionality, retaining the initial 25 principal components. Since these two datasets do not contain user profiles, we used the interaction data from the first week to construct the contextual features of users. Specifically, we used the average vector of all the POI category feature vectors that the user checked in during the first week as the user context feature vector. For the candidate action set at each time step, we selected the ground-truth check-in tag and randomly extracted 24 negative categories of the current step. We used all the data from April 10th, 2012 to February 16th, 2013 (except data from the first week) to construct a data streaming. 

\begin{table}[!htbp]
\centering
\caption{The statistics of the Foursquare NYC \& TKY.}
\vspace{-1ex}
    \label{tab:foursquare data info}
    \begin{tabular}{ccccc}
    \toprule
    Dataset & \#Users & \#POIs & \#POI Categories & \#Check-ins  \\
    \hline
    NYC & 1,083 & 38,333 & 400 & 227,428    \\
    TKY & 2,293 & 61,858 & 385 & 573,703    \\
    \bottomrule
    \end{tabular}
\end{table}

As illustrated in Fig.~\ref{subfig:reward:b} and Fig.~\ref{subfig:reward:c}, similar conclusions can be drawn as in KuaiRec. Additionally, we conducted an analysis of time cost for all algorithms and compared the performance of different estimated rank  ($\tau=1$, $\tau=5$ and w/o Low-Rank) of HyperBandit. The corresponding results are presented in Table~\ref{tab:comparisons of NAR and RT}. Notably, HyperBandit consistently outperformed the baselines in terms of normalized accumulated reward, while the running time of BP (bandit policy) remained acceptable. FactorUCB achieved excellent performance among baselines due to leveraging user adjacency relationships. However, that also lead to a significant time cost as FactorUCB required updating all user parameters at each time step. Furthermore, HyperBandit ($\tau=1$) and HyperBandit ($\tau=5$) reduced the training time of HN by 15.7\%, 8.1\% in KuaiRec and 38.7\%, 30.1\% on Foursquare dataset compared to HyperBandit (w/o Low-Rank), which provided strong evidences of the training efficiency with low-rank factorization.

\subsection{Ablation Experiments}

In this section, we empirically studied the proposed HyperBandit by addressing the following research questions: 

{\raggedright\textbf{RQ1:}} Is HyperBandit efficient enough to meet the real-time requirements of online recommendations?

{\raggedright\textbf{RQ2:}} 
How does the estimated rank $\tau$ of low-rank factorization affect HyperBandit?

{\raggedright\textbf{RQ3:}} What is the impact of key components in HyperBandit on the recommendation performance?

\subsubsection{\textbf{RQ1: Running Time}} 
In streaming recommendation scenarios, running time is another important metric. We reports the running time of bandit policy and the training time of hypernetwork in Table~\ref{tab:comparisons of NAR and RT}. From the results, we conclude that the time cost of HyperBandit was on the order of milliseconds (ms), indicating that HyperBandit met the real-time requirements in streaming recommendations. 
Furthermore, the training time of the hypernetwork exhibits a decreasing trend as the estimated rank $\tau$ decreases from full rank to $1$, validating its efficiency in low-rank updating.

\subsubsection{\textbf{RQ2: Impact of  Estimated Rank $\tau$}}

Fig.~\ref{fig:ranks performance} explored the impact of different estimated rank of low-rank factorization on the performance in terms of normalized accumulated reward and training time. The observations from the experimental results can be summarized as follows: 1). With an increase in the estimated rank $\tau$, our HyperBandit demonstrated an overall improvement in normalized accumulated reward on Foursquare datasets, and the normalized accumulated reward of HyperBandit with ranks ranging from 2 to 14 were nearly equivalent to that of HyperBandit without low-rank factorization. Particularly on KuaiRec dataset, HyperBandit with ranks from 1 to 9 even outperformed the algorithm without low-rank factorization. These results validate the effectiveness of the low-rank factorization approach, which maintains excellent performance. 2). The average epoch, which measures the training time of the hypernetwork, also exhibits an overall upward trend as the estimated rank $\tau$ increases, although it is significantly smaller than that without low-rank factorization. This observation highlights the benefits of efficient training via low-rank factorization as described in Sec.~\ref{sec:HB:Lowrank:training}.


\subsubsection{\textbf{RQ3: Impact of Key Components in HyperBandit.}}
HyperBandit consists of two key components for online updating: one is to update the latent features of items via ridge regression, and the other is to update the parameters of the hypernetwork through gradient descent. 
To investigate the interplay between these two updating components, an ablation experiment was conducted with the following settings:
1). Disable ridge regression updating: The dimension of the latent item features was set to zero.
2). Disable hypernetwork updating: The hypernetwork parameters were frozen to their initial state. 
The results are presented in Table~\ref{tab:comparisons of NAR and RR}. Based on the results, the following conclusions can be drawn:
1). Employing both updating components independently enhances the recommendation performance.
2). Irrespective of whether ridge regression was enabled or disabled, the utilization of the hypernetwork can lead to performance improvements.
\vspace{-.5ex}

                                  

\begin{table}[!h]
    \centering
    \caption{The results of the ablation experiment on key components of HyperBandit. Note that the ridge regression updating in the bandit policy is denoted by ``RR'', and the hypernetwork updating is referred to as ``HN''. The symbol \checkmark~ signifies the inclusion of a particular update process, while the symbol \xmark~ indicates its exclusion. }
    \vspace{-1ex}
    \label{tab:comparisons of NAR and RR}
    \begin{subtable}[ht]{.9\linewidth}
    \centering
    \begin{tabular}{c|c|c|c|c}
        \hline
        \multirow{2}{*}{RR} & \multirow{2}{*}{HN} & \multicolumn{3}{c}{Normalized Accumulated Reward}  \\
        \cline{3-5}
         ~&~& KuaiRec &   NYC   & TKY  \\ 
        \hline \hline
        \xmark      & \xmark      & $0.93\pm0.08$  & $0.87\pm 0.08$   &  $1.14\pm 0.44$      \\
        \checkmark       & \xmark       &$1.09\pm 0.08$  & $5.62\pm 0.52$  &$13.26\pm 0.23$    \\
        \xmark     & \checkmark       & $1.86\pm 0.09$   & $4.90\pm 0.28$  & $10.91\pm 0.35$   \\ 
        \checkmark    & \checkmark      & $\mathbf{3.24\pm 0.11 }$ & $\mathbf{8.27\pm 0.17}$ &  $\mathbf{14.49\pm 0.07}$\\
        \hline
    \end{tabular}
    \label{tab:bt:bandit:SBUCB}
    \end{subtable}
    \vspace{-1ex}
\end{table}


\vspace{-.5ex}
\section{Conclusion}
This paper aims to model the user preference shift in periodic non-stationary streaming recommendation scenarios. 
Specifically, we propose an online learning approach called HyperBandit.
The proposed HyperBandit leverages a hypernetwork to dynamically adjust user preference parameters for estimating time-varying rewards, employs a bandit policy for online recommendation with a regret guarantee, and utilizes a low-rank factorization method to efficiently train the model. 
Experimental results demonstrated the effectiveness and efficiency of HyperBandit in steaming recommendation. The proposed HyperBandit has opened up a promising avenue for advancing controllable online learning.


\balance
\bibliographystyle{ACM-Reference-Format}
\bibliography{sample-base}

\end{document}